%% file: frb20230708a_main.tex
\documentclass[twocolumn]{aastex631}

\usepackage{xcolor}
\newcommand{\changes}[1]{{\textcolor{black}{#1}}}

\newcommand{\ourfrb}{FRB\,20230708A}
\newcommand{\ourhost}{J201227.73-552122.72}

\newcommand{\nfurby}{12}

\newcommand{\dmunits}{\ensuremath{{\rm pc \, cm^{-3}}}}

\newcommand{\fastfrb}{FRB\,20190520B}

\newcommand{\dmmwism}{\ensuremath{\mathrm{DM}_\mathrm{MW}^\mathrm{ISM}}}

\newcommand{\halpha}{\ensuremath{\mathrm{H}\alpha}}
\newcommand{\hbeta}{\ensuremath{\mathrm{H}\beta}}

\newcommand{\pox}{\ensuremath{P(O|x)}}

\newcommand{\xshoot}{X-Shooter}
\newcommand{\xshootslit}{X-Shooter slit}

\newcommand{\lstar}{\ensuremath{L_*}}

\newcommand{\rmag}{\ensuremath{22.53\pm0.02}}
\newcommand{\ourz}{\ensuremath{0.1050\pm0.0001}}
\newcommand{\ourZ}{\ensuremath{12+\log(\text{O}/\text{H})\sim(\changes{8.0}-8.3)}}
\newcommand{\ourL}{\ensuremath{1.6\times 10^8L_\odot}}

\usepackage{CJKutf8}

\begin{document}

\title{The Low-mass Dwarf Host Galaxy of Nonrepeating FRB 20230708A}

\input{affiliations}

\correspondingauthor{August~R.~Muller}
\email{armuller23@gmail.com}

\author[0000-0002-2849-6955]{August~R.~Muller}
\MMO
\AEI
\Glasgow

\author[0000-0002-5025-4645]{Alexa~C.~Gordon}
\NU
\CIERA

\author[0000-0003-4501-8100]{Stuart D. Ryder}
\MacquarieMathPhys
\MacquarieSpace

\author[0009-0002-5700-7978]{Alexandra G. Mannings}
\UCSC

\author[0000-0002-7738-6875]{J.~Xavier~Prochaska}
\UCSC
\IPMU
\NAOJ

\author[0000-0003-2149-0363]{Keith~W.~Bannister}
\CSIRO
\Sydney

\author[0000-0002-2864-4110]{A.~Bera}
\Curtin

\author[0000-0003-3460-506X]{Shivani Bhandari}
\CSIRO
\SKAO

\author[0000-0002-8383-5059]{N.~D.~R.~Bhat}
\Curtin

\author[0000-0001-9434-3837]{Adam~T.~Deller}
\Swinburne
\OzGrav

\author[0000-0002-7374-935X]{Wen-fai Fong}
\NU
\CIERA

\author[0000-0002-5067-8894]{Marcin Glowacki}
\Edinburgh
\CapeTown

\author[0000-0001-9817-4938]{Vivek Gupta}
\CSIRO

\author[0000-0003-4193-6158]{J.~N.~Jahns-Schindler}
\Swinburne

\author[0000-0002-6437-6176]{C.~W.~James}
\Curtin

\author[0000-0003-2973-0472]{Regina~A.~Jorgenson}
\MMO
\HumboldtCA

\author[0000-0003-1483-0147]{Lachlan Marnoch}
\MacquarieMathPhys
\MacquarieSpace
\CSIRO

\author[0000-0002-7285-6348]{R.~M.~Shannon}
\Swinburne

\author[0000-0002-1883-4252]{Nicolas Tejos}
\Valparaiso

\author[0000-0002-2346-6853]{P.~A.~Uttarkar}
\Swinburne

\author[0000-0003-0203-1196]{Yuanming Wang}
\Swinburne
\OzGrav

\author[0000-0002-2066-9823]{Ziteng Wang \begin{CJK*}{UTF8}{gbsn}(王子腾)\end{CJK*}}
\Curtin

\begin{abstract}

We present Very Large Telescope/X-Shooter spectroscopy
for the host galaxies of \nfurby\ fast radio bursts
(FRBs) detected by the Australian SKA Pathfinder observed through the ``Fast and Unbiased FRB Host Galaxy (FURBY)" Large Programme at the European Southern Observatory, which imposes strict selection
criteria on the included FRBs and their host galaxies to produce
a homogeneous and well-defined sample.  
We describe the data reduction and analysis of these spectra and report their redshifts, line-emission fluxes, and derived host properties. 
From the present sample, this paper focuses
on the faint host of \ourfrb\
($m_R = \rmag$) identified at low redshift
($z=0.1050$). This indicates an intrinsically very
low-luminosity galaxy ($L \approx 10^8 L_\odot$),
making it the lowest-luminosity nonrepeating FRB host to date by a factor of $\sim 3$ and slightly dimmer than the lowest-luminosity host for repeating FRBs. 
Our spectral energy distribution fitting analysis reveals a low stellar mass
($M_* \approx 10^{8.0} M_\odot$), low star
formation rate 
(${\rm SFR} \approx 0.04 M_\odot \, \rm yr^{-1}$),
and very low metallicity (\ourZ),
distinct from the more massive galaxies (log($M/M_\odot$) $\sim$ 10) that are commonly identified for nonrepeating FRBs.
Its discovery demonstrates that FRBs can arise
in the faintest, metal-poor galaxies of the Universe.  
In turn, this suggests that at
least one FRB progenitor channel must include stars 
(or their remnants) created in very low metallicity
environments.  This indicates better prospects
for detecting FRBs from the high-$z$ Universe
where young, low-mass galaxies proliferate.

\end{abstract}


\section{Introduction} \label{sec:intro}

The transient sky is now rich with a variety of phenomena that
change in location and/or luminosity over human timescales,
i.e., seconds to years. This includes exploding stars,
active galactic nuclei (AGN), near-Earth objects, and pulsars.
Their study drives our understanding of the astrophysics
of compact objects and accretion. Our knowledge of the transient sky will only continue to improve and diversify with the onset of new facilities such as the Vera C. Rubin Observatory \citep{LSST-system}. 

One of the most recent classes of transients are fast
radio bursts (FRBs): bright, millisecond-duration pulses of radio emission generally detected at frequencies $\nu \approx$ 0.4--1.5 GHz. Though the first FRB was reported in 2007 \citep{Lorimer+2007}, the first FRB host galaxy was confidently established in 2017, confirming that FRBs are of extragalactic origin \citep{clw+17, mph+17, tbc+17}. Some repeating FRB sources have been seen to produce several to hundreds of bursts \citep{CHIMErepeaters,HyperactiveRepeater}, while the majority of sources are associated with only one FRB detection \citep{CHIMEcat1,CHIMECat2}. 

Despite \changes{thousands of} published FRB detections to date (e.g., \citealt{CHIMECat2}), including $\sim$100 with confident host associations (e.g., \citealt{Bannister+19,Ravi+19,bhandari2020,Marcote_etal_2020,Gordon2023,Law+24,Sharma2024,CHIME+25,shannon2025,Pastor-Marazuela+26}), FRB progenitors and their emission mechanisms are still not well understood \citep{Zhang2023}. 
This has led the community to pursue a variety of approaches
to rule out candidate models, including via FRB pulse characteristics or population demographics (e.g., \citealt{Pleunis21,CHIMErepeaters,Curtin+24,Scott2025}).
For the subset of FRBs localized with high probability to a host galaxy \citep{path}, 
such studies have also investigated the source 
location within the galaxy \citep[e.g.,][]{Bassa+17,Mannings2021,Tendulkar+21,Dong+24,Gordon+25}
and host galaxy demographics 
\citep[e.g.,][]{bhandari2020,heintz2020,Gordon2023,Sharma2024}.
To date, these studies have ruled out AGN as the leading source and have identified emission line (i.e., star-forming) galaxies as the predominant \citep[but not sole; see][]{Eftekhari+25} 
hosts of FRBs, suggesting a young progenitor population \citep{Eftekhari2023}.
The coincidence of FRB-like emission with the position of Galactic magnetar SGR 1935+2154 provided strong support for a magnetar progenitor channel \citep{magnetar_det}, in good agreement with the fact that nearly all FRB host galaxies exhibit active star formation \citep{Gordon2023,Sharma2024}.
However, existing host galaxy samples have been drawn from a 
heterogeneous set of radio surveys and host follow-up
strategies. As such, conclusions drawn to date likely
suffer from selection biases that may complicate 
interpretations and constraints on FRB progenitors (e.g., see the competing conclusions of \citealt{Sharma2024} \changes{and} \citealt{horowicz2025}).

With host distributions and demographics as a primary
motivation, we launched the 
\textbf{F}ast and \textbf{U}nbiased F\textbf{RB} Host Galax\textbf{y} (FURBY; Large Programme 108.21ZF, PI Shannon) 
program on the European Southern Observatory's Very Large Telescope (VLT) to 
uniformly follow up FRBs from the Commensal Real-time ASKAP Fast Transients (CRAFT) survey \citep{CRAFT,shannon2025} on the Australian SKA Pathfinder \citep[ASKAP;][]{ASKAP}.
By adopting strict selection criteria for FRB
inclusion and follow-up procedures (see Section \ref{sec:obs} for full details), 
we present an initial homogeneous sample of \nfurby\ 
FRBs with very high posterior probability host
associations $\pox > 0.99$.

This paper reports the discovery and analysis
of the host galaxy of as-yet-nonrepeating \ourfrb, an FRB with unusual burst properties \citep{Dial2025} detected by ASKAP/CRAFT \citep{shannon2025} and observed
as part of the FURBY program.
Our deep imaging and follow-up spectroscopy reveal that
this is the faintest known host from a nonrepeating FRB to date.
Notably, given that it was selected from a parent sample of only 12~FRBs, this indicates low-luminosity
hosts \changes{make up a non-negligible fraction of FRB host galaxies}, even among the apparently nonrepeating population. This discovery is timely in that it may contradict recent
conclusions on a bias against low-metallicity hosts
from a heterogeneous sample \citep{Sharma2024}.
Furthermore, it also goes against the developing convention that such faint hosts are exclusively associated
with repeating FRBs (e.g., \citealt{tbc+17,Hewitt2024}).

In this work, we present the first FURBY sample and provide spectroscopic results for 12 FURBY host galaxies, with particular emphasis on the notable host galaxy of \ourfrb. Section \ref{sec:obs} describes the FURBY program and details of the associated spectroscopic observations. Section \ref{sec:results} presents a detailed analysis of the host galaxy of \ourfrb, including its gas emission properties (Section \ref{sec:gas_emission}), its mass and star formation history (SFH) through the use of spectral energy distribution (SED) modeling (Section \ref{sec:prospector}), its luminosity in the context of other known FRB hosts (Section \ref{sec:lum})\changes{, and constraints on an associated persistent radio source (PRS; Section \ref{sec:prs})}. Finally, we discuss the implications of this host galaxy in Section \ref{sec:discussion}.
We use AB magnitudes and WMAP9 cosmology
throughout \citep{wmap9}.

\section{The FURBY Sample and Observations} \label{sec:obs}

FRBs detected by ASKAP/CRAFT starting in January~2022 are eligible for observation in the FURBY survey.
To qualify, the candidate must
pass a series of minimum observation criteria, designed to produce a \changes{uniform sample of FRB host galaxies with well-understood selection effects. }

\changes{
Radio selection effects are a function of observation frequency, beam pattern, maximum searched dispersion measure (DM), and integration time. These selection effects are modeled by \citet{shannon2025} in the case of the incoherent sum observations that detected all but two of the FRBs analyzed here. The remaining two, FRBs~20231230D and 20240117B, were detected by ASKAP's coherent CRACO system, as modelled by \citet{2025PASA...42....5W}. These effects should be accounted for when modeling the DM-z distribution of FRB hosts \citep{2022MNRAS.509.4775J}. Importantly, all FRBs are detected in a blind survey (as opposed to targeted follow-up of a previously identified repeating FRB), and the gain in signal-to-noise in offline analysis \citep{2023A&C....4400724S} means that even FRBs which are marginally detected in the real-time system have arcsecond-scale radio localizations. }

\changes{To minimize optical biases in the association
of each FRB to a host galaxy, we then follow-up the sample with} a uniform suite of instruments to a common depth (i.e., a magnitude limited sample). The first set of FURBY criteria based on the 
CRAFT localization are imposed before imaging: 

\begin{enumerate}
    \item Galactic reddening must be $E(B-V) < 0.1 \, \rm mag$, 
    as given by the dust maps from \cite{SchlaflyFinkbeiner2011}.
    \item The Galactic dispersion measure contribution must be
    $\dmmwism< 100\,\dmunits$,
    as computed using the model of \cite{ne2001}.
    \item There must be no nearby bright star that may interfere with host galaxy identification and spectroscopy\footnote{The radius affected by a bright star as a function of its magnitude is given as $\rm r \approx 1.8 + 0.4 × \exp[(20 - \textit{R})/2.05]\, \rm arcsec$ (see criterion 6 in \citealt{hjorth+2012}).}.
    \item The total $1\sigma$ uncertainty 
    of the major axis of the FRB localization 
    must be less than $0\farcs7$.
\end{enumerate} 
These criteria mitigate against Galactic extinction
and reduce ambiguity with host associations
\citep{tb2017,path}.
 
FURBY candidates passing these criteria were observed in the $R$ and $K_{\rm s}$ bands using the VLT FOcal Reducer/low dispersion Spectrograph 2 (FORS2) and High Acuity Wide-field \textit{K}-band Imager (HAWK-I; with the Ground-layer Adaptive optics Assisted by Laser system), respectively. The imaging procedure, data reduction, astrometric and flux calibration, and host galaxy photometry followed the process laid out in \citet{Marnoch2023}
with typical $5\sigma$ depths of $R\sim26.5$\,mag and $K_{\rm s}\sim23.5$\,mag 
in a $1''$~diameter circular aperture.

The $R$-band images were then used to perform a Probabilistic Association of Transients to their Hosts (PATH; \citealt{path}) analysis to determine the most likely host in each case. We use the standard priors commonly adopted in the literature (e.g., \citealt{Sharma2024,Hewitt2024}), which include the inverse magnitude candidate prior and the exponential profile offset prior with a scale
factor of 0.5 \citep{shannon2025} truncated at six effective radii. Two further criteria were then imposed based on the imaging results to trigger spectroscopic follow up:

\begin{enumerate}
  \setcounter{enumi}{4}
  \item The PATH posterior probability \pox\ 
  for the highest likelihood host
  must exceed 0.4 (in practice, all FURBY hosts passing this criterion had $\pox> 0.99$).
  \item The highest likelihood host must be brighter than $m_{R} = 24$~mag such that spectroscopy had a reasonable
  chance of successfully yielding a redshift with ground\changes{-}based facilities. 
\end{enumerate}

FRB hosts meeting these additional criteria were then observed using the \xshoot\ spectrograph covering 0.3--2.5$\,\mu$m. In practice, all FRBs that passed criteria 1$-$4 also passed criteria 5 and 6 and were observed with \xshoot. The broad spectral coverage offered by \xshoot\ is advantageous for assessing a wide range of galaxy redshifts
(the FURBY sample spans $z \sim$~[0.04,1.02]), particularly when key diagnostic lines span both the optical and near-infrared (NIR) regimes that would otherwise require the use of two different spectrographs (and potentially result in two mismatched slit positions). 
All spectroscopic observations used a $1\farcs3 \times 11\arcsec$ slit in the ultraviolet-blue (UVB) arm and $1\farcs2 \times 11\arcsec$ slits in the VIS and NIR arms, yielding a spectral resolving power of 4100, 6500, and 4300 in the UVB, VIS, and NIR arms, respectively\footnote{In the case of \changes{the $z\sim1$} FRB\,20220610A, narrower slits of $1\farcs0$ and $0\farcs9$ were employed, resulting in larger resolving powers of 5400, 8900, and 5600.}. 
Exposure times were adjusted based on the brightness of the host and are listed in Table~\ref{tab:observations}. Further observational details for all FRB hosts in our sample are also given in Appendix~\ref{ap:obs_details}. 
In general, host galaxies brighter than $m_{R} = 23$ were observed in STARE mode fixed at the center of the slit\footnote{In the case of the edge-on host for FRB~20240201A which filled the entire slit, a matching observation of blank sky $20\arcsec$ to the South was also taken.}, while fainter hosts used the NOD mode cycling the target $\pm3\arcsec$ along the slit's longer axis. Observations of white dwarf stars and B9~V stars were used for relative spectrophotometric calibration
and telluric feature removal, respectively.

We used the spectroscopic reduction software \texttt{PypeIt}
\citep{pypeit:joss_pub,pypeit:zenodo} to process the \xshoot\ data and prepare them for release. 
\xshoot\ consists of three echelle spectrographs. For reduction of each of the NIR exposures, we apply calibrations, which includes slit tracing, flat-fielding, and dark subtraction. We do not apply bias subtraction, as the NIR detector is sensitive to dark current and the master dark frame captures the bias of the detector as well. We follow a similar scheme for the reduction of the VIS and UVB exposures; however, we apply bias subtraction (without dark subtraction). Wavelength calibrations are applied for the UVB and VIS arms using vacuum reference wavelengths. We then subtract skylines from each calibrated exposure.
Once these steps, along with flux calibration, have been completed, we co-add the exposures and extract a final one-dimensional spectrum for each host. See Section \ref{sec:gas_emission} for the resulting spectrum of the host galaxy associated with \ourfrb\ and Appendix \ref{ap:obs_details} for all other spectra included in this work. 

\begin{figure}
    \centering
    \includegraphics[width=0.45\textwidth]{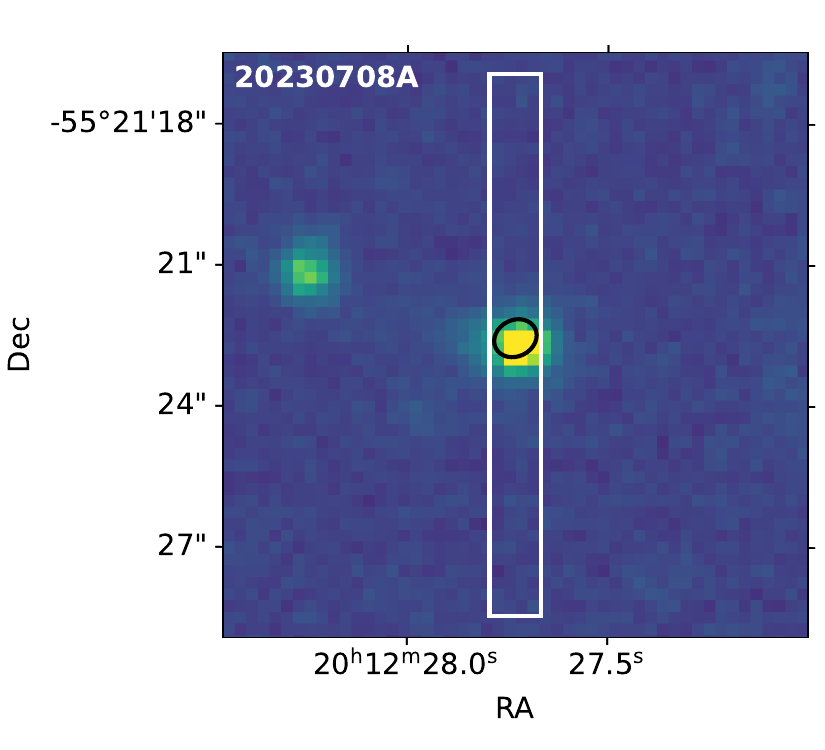}
    \caption{VLT/FORS2 $R$-band image of the host galaxy associated with \ourfrb\ (center of image). The VLT/\xshoot\ slit outline is shown in white, and the FRB localization ellipse is shown in black.}
    \label{fig:slit_230708}
\end{figure}

In Figure~\ref{fig:slit_230708}, we present the FORS2 $R$-band image of the host galaxy of \ourfrb, 
with outlines denoting the \xshoot\ slit position and FRB localization region.
We present the images and slit positions of 11 additional hosts observed through this program using FORS2 and \xshoot\ in Appendix~\ref{ap:obs_details}, representing all FURBY hosts spectroscopically observed through May 15, 2024. 
This FURBY sample marks a uniquely homogeneous 
set of FRB host galaxies spanning substantial cosmic time ($0\lesssim z \lesssim 1$).

\section{Characterizing the Host Galaxy \changes{and Environment} of \ourfrb} \label{sec:results}

\subsection{Emission Lines} \label{sec:gas_emission}

\begin{figure*}
    \centering
    \includegraphics[width=\textwidth]{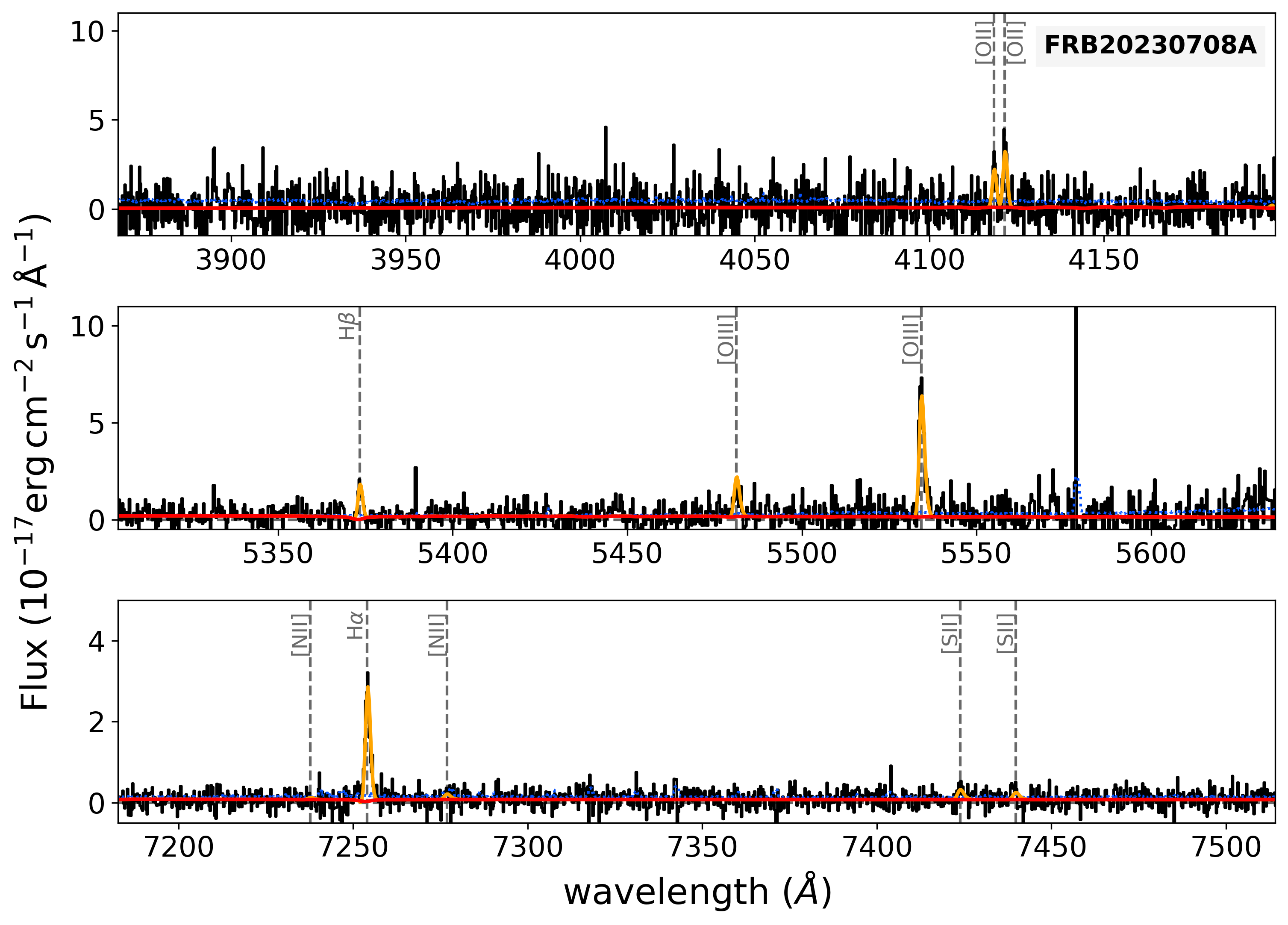}
    \caption{VLT/\xshoot\ spectrum of the \ourfrb\ host galaxy showing prominent H$\alpha$, H$\beta$, [\ion{O}{3}], and [\ion{O}{2}] emission. There is a notable lack of clear [\ion{N}{2}] emission features. The pPXF gas emission (orange) and stellar continuum (red) best fits are shown. Spectral error is shown in blue.}
    \label{fig:1D-230708}
\end{figure*}

As described in \cite{Dial2025}, \ourfrb\ was
associated at high confidence to the galaxy
\ourhost\ using the PATH formalism \citep{path}.
Following spectroscopic reduction, we use the Penalized Pixel-Fitting (\texttt{pPXF}) package \citep[][including stellar emission templates from \citealt{E-MILES}]{ppxf} to fit and measure emission lines, which are used to determine its redshift. The resulting fit is shown in Figure~\ref{fig:1D-230708}. We repeat this process for all other spectra in the sample and present the measured gas emission fluxes from key emission lines in Appendix~\ref{ap:fluxes}. While no stellar continuum is detected from the host of \ourfrb,\ there is confident detection of H$\alpha$, H$\beta$, [\ion{O}{2}]$\lambda\lambda3726,3729$, and [\ion{O}{3}]$\lambda\lambda4959,5007$ emission, as shown in Figure~\ref{fig:1D-230708}. Notably, the [\ion{N}{2}]$\, \lambda\lambda6548,6583$ emission line doublet is not detected. 
Based on the locations of H$\alpha$, H$\beta$, and [\ion{O}{3}], we determine a redshift for this host galaxy of $z=\ourz$. 

\begin{figure}
    \centering
    \includegraphics[width=0.5\textwidth]{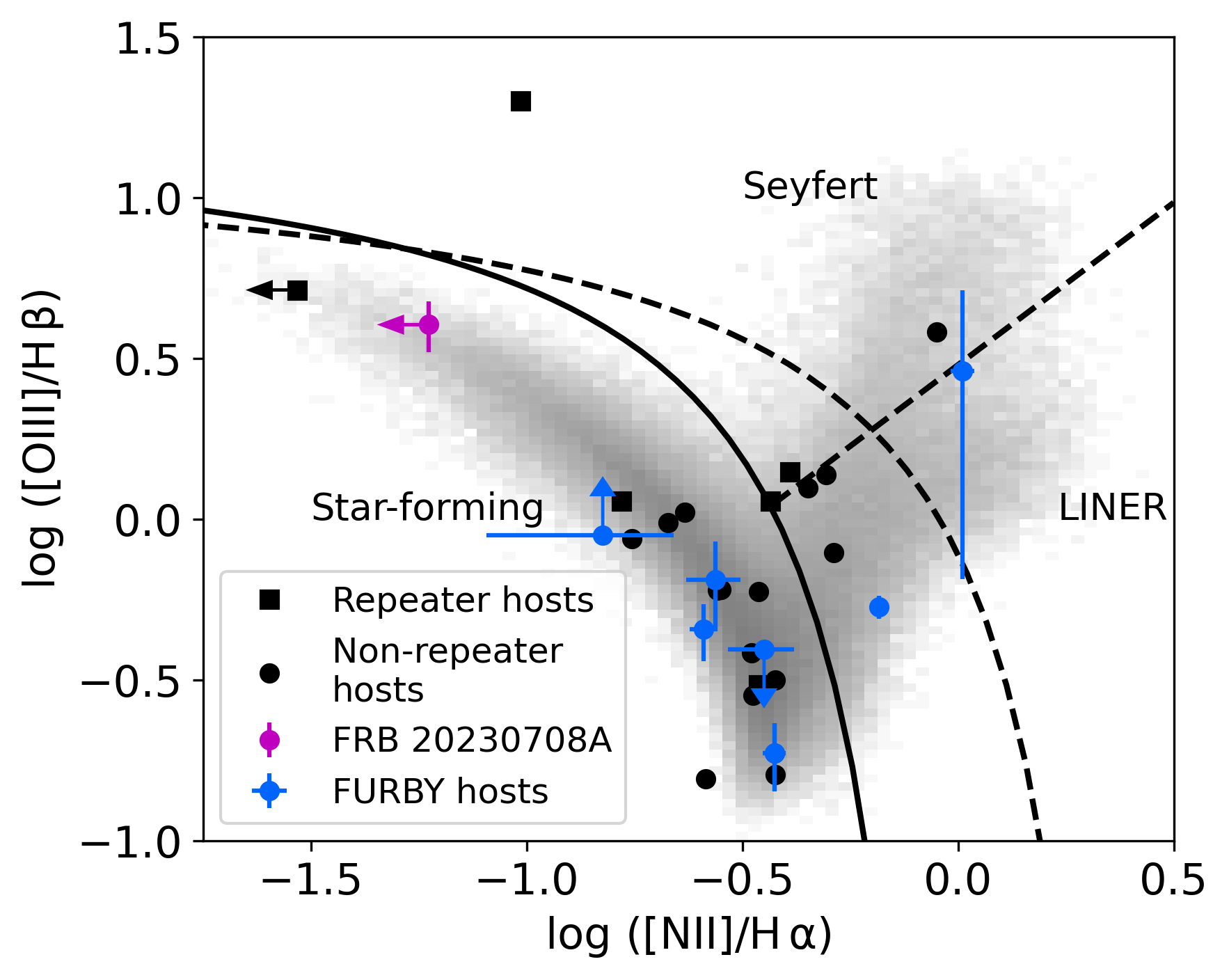}
    \caption{A BPT classification diagram indicating the dominant ionization mechanisms of FRB host galaxies. Previously published FRB host galaxies (see \citealt{Eftekhari2023}) are shown in black; all hosts from the FURBY survey \changes{presented here} are included in blue. The host galaxy of \ourfrb~is emphasized in magenta. A background sample of all galaxies in the SDSS catalog is \changes{shown in} gray.}
    \label{fig:bpt}
\end{figure}

We next use the observed gas emission \changes{line} fluxes (or \changes{upper} limits) to place this host on a Baldwin-Phillips-Terlevich (BPT) diagram, shown in Figure~\ref{fig:bpt} (\citealt{BPT}; including classifications specified in \citealt{CidFernandes2010,Kauffmann2003,Kewley2001}). The host of \ourfrb\ clearly lies within the regime where star formation is the dominant ionization mechanism.

Based on the measured gas emission lines H$\alpha$, H$\beta$, [\ion{O}{3}], and our upper limits on [\ion{N}{2}], we find a O3N2 limit of $\geq0.6$ and an N2 limit of $\leq-0.4$. Using calibrators from \citet{O3N2}, these bounds yield metallicity limits of $\changes{8.0}<12 + \log({\rm O/H}) <8.3$, respectively, indicating a relatively low total metallicity for the \ourfrb~host.

\subsection{SED Modeling} \label{sec:prospector}

To derive the stellar population properties of the host galaxy, we use the Bayesian inference code \texttt{Prospector} \citep{Johnson+21}. \texttt{Prospector} jointly fits the provided photometry and spectroscopy to SED models generated with \texttt{python-fsps} \citep{Conroy2009, Conroy2010}. We use the \citet{Kroupa01} initial mass function, \citet{KriekandConroy13} dust attenuation curve, an eight bin \texttt{continuity} non-parametric SFH \citep{Leja2019}, and require adherence to the \citet{Gallazzi2005} mass-metallicity relation. Further, we employ a model to normalize the photometry to the spectroscopy using a 12th order Chebyshev polynomial, a spectral smoothing model, a jitter model to adjust for noise in the observed spectrum, a pixel outlier model to marginalize over poorly modeled noise, and finally, a model to marginalize over the spectral emission lines. For further details on these priors and their allowed ranges, see \citet{Gordon2023}. Once the data have been used to constrain the priors, we use the nested sampling routine \texttt{Dynesty} \citep{Speagle2020} to sample the posterior distributions.

To supplement the VLT photometry for the \texttt{Prospector} modeling, we use $griz$ band photometry from the Dark Energy Survey \citep{DES}. All photometry measurements are corrected for Galactic extinction using the \cite{Fitzpatrick:2007} extinction law. We jointly fit the photometry with our spectroscopy, using the VIS arm portion of the \xshoot\ spectrum, which we similarly correct for extinction with the \cite{Fitzpatrick:2007} law.

The best-fit model reveals a very low-mass (log(M$_*$/M$_{\odot}$) = $7.97^{+0.09}_{-0.08}$) dwarf galaxy with a current star formation rate of $0.04^{+0.02}_{-0.01}$\,M$_{\odot}$~yr$^{-1}$ and mass-weighted age of $5.82^{+0.94}_{-1.25}$~Gyr (all reported values represent the median and 68\% confidence interval). Per the mass-doubling number criterion of \citet{Tacchella+22} to assess degree of star formation, the specific star formation rate of log(${\rm sSFR}_{\rm 0-100 Myr}$) = $-9.62^{+0.27}_{-0.24}$ indicates that the host of \ourfrb\ is actively star-forming. This is in agreement with its position on the BPT diagram.

\begin{figure}
    \centering
    \includegraphics[width=0.465\textwidth]{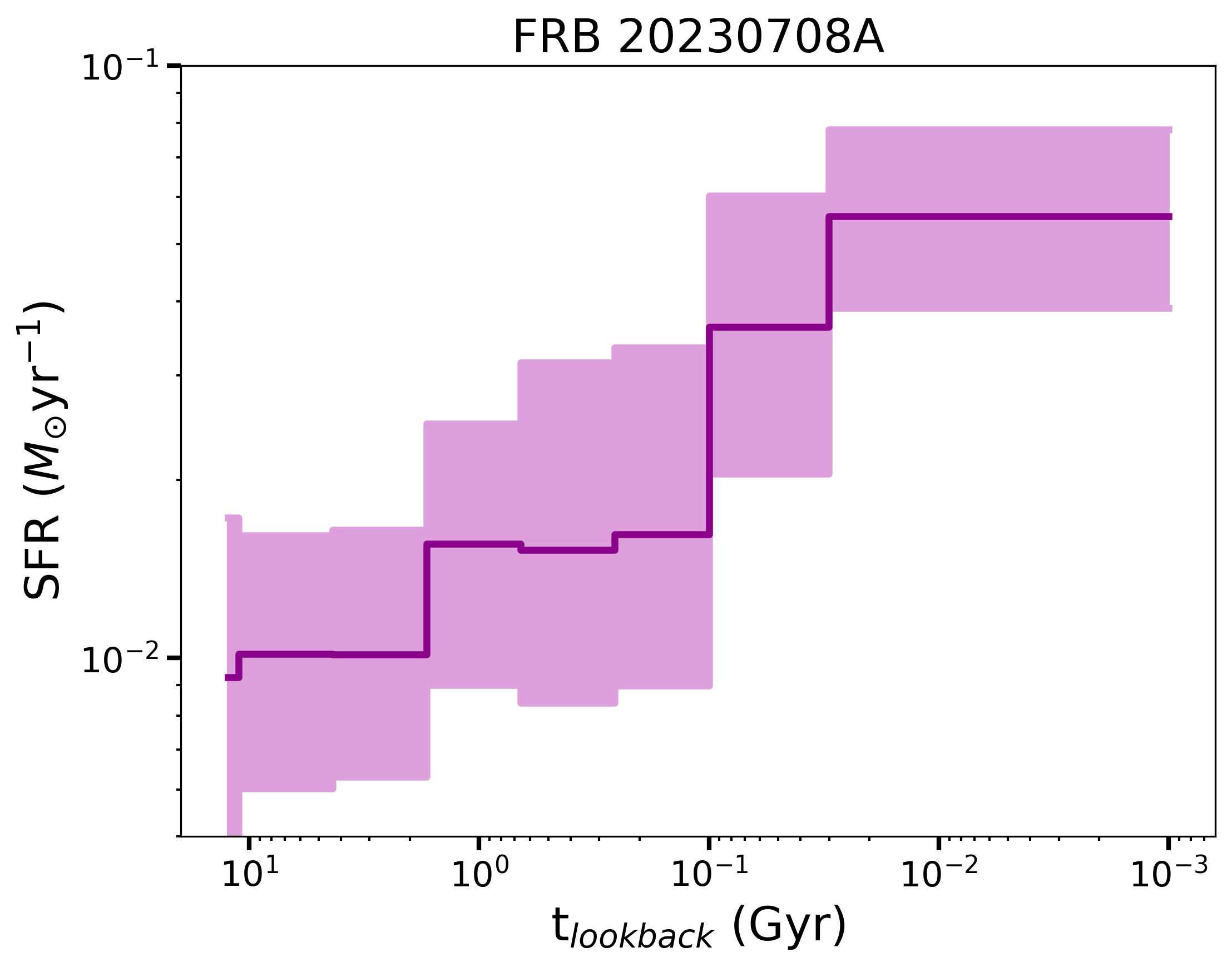}
    \caption{The SFH of \ourfrb. The increase in SFR towards present day suggests that this host is actively building up its mass.}
    \label{fig:SFH}
\end{figure}

In Figure~\ref{fig:SFH}, we present the SFH, which shows a steady increase in star formation towards the present day. This indicates that the galaxy may still be in the process of building up its mass, commensurate with its current low mass. Interestingly, this rising SFH behavior is consistent with the hosts of several FRBs that are known to repeat \citep{Gordon2023}. While the age and specific star formation rate of \ourfrb's host are representative of the larger FRB host population, it is significantly less massive than any other known nonrepeating FRB host galaxy. 

\subsection{Luminosity Comparison} \label{sec:lum}

\begin{figure}
    \centering
    \includegraphics[width=0.45\textwidth]{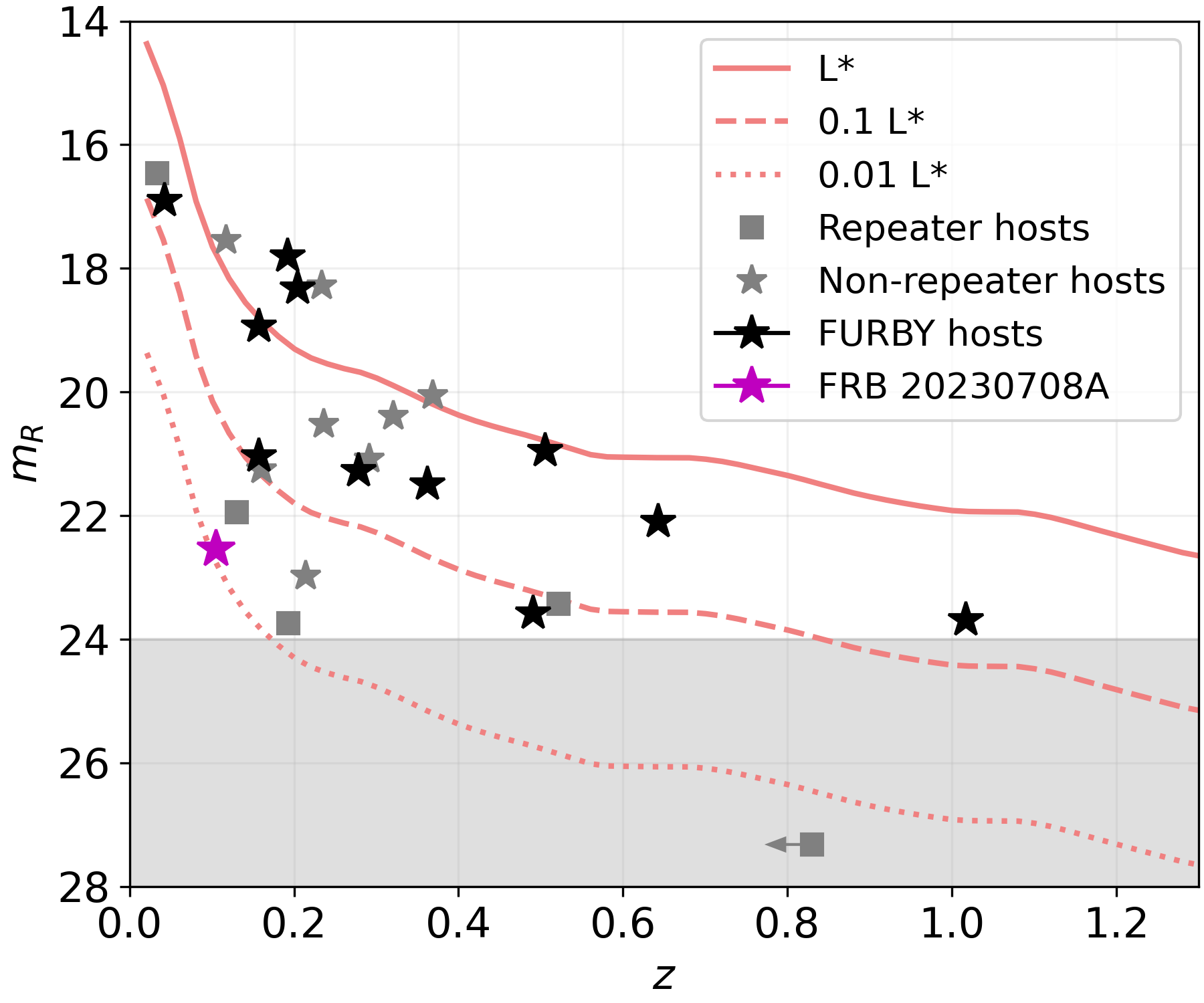}
    \caption{Apparent magnitude versus redshift for all FRB hosts in \changes{\citet{Bhandari2023}, \citet{Hewitt2024}, \citet{Bhardwaj2025}, and} \citet{FRB_Repo}
    with $\pox > 0.9$ and a published $m_R$ or $m_r$ value. 
    FURBY hosts are shown as stars (FRB 20230708A is labeled accordingly; all FURBYs correspond to non-repeating FRBs). Other public FRB host galaxies are shown in gray (shape depends on repeater status). The shaded region indicates the portion of parameter space excluded from FURBY due to the imposed magnitude cut. \changes{The \citet{Hewitt2024} host is shown as a limit at the published photometric $z_{\text{max}}$ as no spectroscopic redshift is available.}}
    \label{fig:Lstar}
\end{figure}

In order to place the host of \ourfrb\ in the broader context of published FRB hosts, we show the redshift-evolving luminosity function of galaxies \citep[parameterized by $\lstar$, the characteristic luminosity scale of the Schechter luminosity function, from][]{Schechter1976}.
In Figure~\ref{fig:Lstar}, we plot the $R$-band magnitude versus redshift for all FURBYs in the current sample, as well as a broader sample of published FRB hosts\footnote{Available from \url{frb-hosts.org}.} \citep[in this case hosts are plotted with either $R$-band or $r$-band magnitudes as available;][]{FRB_Repo}. We also plot curves corresponding to $\lstar$, $0.1\lstar$, and $0.01\lstar$ as a function of redshift, each corresponding to the relevant luminosity value in the rest-frame \citep{Brown+01,Wolf+03,Willmer+06,Reddy+09,Finkelstein+15,heintz2020}. We find that the \ourfrb\ host has a very low luminosity, $L = \ourL$, consistent with $0.01\lstar$. 

\changes{\subsection{Constraints on a PRS} \label{sec:prs}}

\changes{Other FRBs associated with low-mass host galaxies have coincident PRSs \cite[][]{clw+17,Niu22}, compact radio continuum sources of unknown nature but postulated to be AGN or pulsar wind nebulae \cite[][]{clw+17}.  Motivated by this, we observed the location of \ourfrb\ with the Australia Telescope Compact Array (ATCA) to search for any PRS coincident with the FRB position. The source was observed in four epochs in four bands centered at 5500, 7500, 16,700, and 21,200\,MHz, each with a bandwidth of 2048\,MHz and a spectral resolution of 1\,MHz (project code C3347; PI Bhandari). The observation used the Compact Array Broad-band Backend \citep[][]{Wilson2011MNRAS.416..832W} while ATCA was in the 6A configuration\footnote{\url{https://www.narrabri.atnf.csiro.au/operations/array_configurations/configurations.html}}. We used PKS 1934$-$638 as the primary bandpass calibrator and PKS 1941$-$554 as the secondary phase calibrator.}

\changes{We flagged and calibrated the data with Miriad \citep{Sault1995ASPC...77..433S} and imaged the data with CASA \citep{CASATeam2022PASP..134k4501C}. For the first observation on March 3, 2024 we measured a faint $4\sigma$ source at the FRB position at 7500 MHz, with a flux density of $25\pm7\,\mu$Jy\,beam$^{-1}$. However, the two follow-up observations on August 30, 2024 and September 25, 2024 did not detect this faint source, suggesting that the detection on March 3 was likely a noise outlier. We did not detect any emission coincident with the FRB position in other frequency bands. The resulting $3\sigma$ upper limits are 22\,$\mu$Jy\,beam$^{-1}$ at 5500 MHz, 20\,$\mu$Jy\,beam$^{-1}$ at 7500 MHz (based on the follow-up observation on August 30, 2024), 35\,$\mu$Jy\,beam$^{-1}$ at 16,700 MHz, and 73\,$\mu$Jy\,beam$^{-1}$ at 21,200 MHz. 
With the known luminosity distance of 490\,Mpc, the corresponding $3\sigma$ limit of spectral radio luminosity is $L_\mathrm{7500\,MHz}=5.2\times 10^{27}$\,erg\,s$^{-1}$\,Hz$^{-1}$, which is the deepest limit from our observations. 
}

\section{Discussion} \label{sec:discussion}

Using the observations \changes{presented here}, we find that the host galaxy of \ourfrb\ represents the lowest-luminosity galaxy associated with a non-repeating FRB to date. Based on its apparent magnitude ($m_R = \rmag~\rm mag$) and redshift, the host galaxy has an $R$-band luminosity of $\ourL$, making it a faint dwarf galaxy ($<10^9 L_{\odot}$)
with $L \approx 0.01 \lstar$ at $z \sim 0$ (Figure~\ref{fig:Lstar}).
Indeed, we find it has the lowest intrinsic luminosity of any known host with spectroscopic redshift confirmation
(see \citealt{Hewitt2024} for a limit \changes{based on photometric observations}). Its luminosity is lower than the dimmest previously reported host of a non-repeating FRB by a factor of $\sim 3$ \citep{Bhandari2023}. \changes{While the faintest published hosts outside of this work are predominantly associated with repeating FRBs, the low luminosity of this host indicates that such galaxies should not be associated exclusively with the repeating population.}

Furthermore, this host is classified as a star-forming dwarf galaxy with very low total mass (log($M/M_\odot$) = 7.97) and metallicity (\ourZ). This galaxy shows similarities to the dwarf host galaxies of \changes{repeating}
FRB\,20121102A \citep{tbc+17} and \fastfrb\ \citep{fast}. 
The localization of these repeating FRBs to low luminosity dwarf galaxies has challenged formation mechanisms in which FRB sources \changes{must} track stellar mass or star formation, instead pointing toward the existence of \changes{other} progenitor channels.
Similarly, the host of \ourfrb\ contrasts the growing sample of FRBs localized to more massive and luminous galaxies with relatively older stellar populations and higher metallicities 
\citep[e.g.,][]{Gordon2023,Sharma2024}. 

Its discovery, therefore, suggests a greater diversity of host galaxy environments and \changes{lends support to the growing evidence for} multiple FRB progenitors and/or multiple formation pathways \changes{\citep[e.g.,][]{Kirsten22,Eftekhari+25, Gordon+25,Shah+25}}, particularly including those which are consistent with the lowest end of the galaxy mass distribution. It also implies the existence of progenitors from
metal-poor stars and/or their remnants.
As a result, it \changes{provides compelling support for the idea} that FRBs may occur in the young, metal-poor galaxies of the high-$z$ universe\changes{, as suggested in \citet{Beniamini+2021}}.

Despite similarities between this host galaxy and the hosts of FRB\,20121102A and \fastfrb, the radio pulse of \ourfrb\ is markedly 
different than the repeating FRBs associated \changes{with} dwarf
galaxies, indicating potential differences in both the burst progenitor and its immediate environment. The burst is broad and comprised of numerous discrete components \cite[][]{Dial2025}. The components themselves are broad-band and do not show the ``sad trombone" morphology archetypal of repeating FRBs, including FRB\,20121102A.
\changes{The scattering time of \ourfrb, $0.14\pm0.02$\,ms at a radio frequency of 920 MHz \citep{Dial2025}, is consistent with the low values estimated for FRB\,20121102A \citep[e.g.][and references therein]{Josephy19}, but is more than two orders of magnitude lower than that seen for \fastfrb\ \citep{Ocker22}.} 

\changes{Similarly, the inferred contribution to burst DM from the burst's host galaxy and local environment is considerably lower for \ourfrb\ (where the observed dispersion measure of 411.5 pc cm$^{-3}$ leaves around 200 pc cm$^{-3}$ after the Milky Way and mean intergalactic medium contributions are subtracted) than \fastfrb\ (300-900 pc cm$^{-3}$; \citealp{Niu22,Lee23}), but is comparable to the $\sim$150--200 pc cm$^{-3}$ for FRB\,20121102A \citep{tbc+17}.}
Further distinctions can be found in the burst polarimetry: unlike the vast majority of bursts from repeating FRBs, \ourfrb\ shows significant circular polarization and clear variation in the linear polarization position angle. 
This indicates that there are variations
in the mechanism that produced the FRBs.

Other pulse observables suggest that \ourfrb\ originated in an environment different \changes{from} that of active repeating FRBs such as FRB~20121102A.
The burst has a small rotation measure \cite[RM $=6.90\pm0.04$\,rad\,m$^{-2}$,][]{Dial2025} and shows no evidence for frequency-dependent depolarization \cite[][]{2025arXiv250319749U}.
This is in contrast to the large rotation measure magnitudes observed in repeating sources\changes{: for example, FRB~20121102A has a rotation measure of $\mathcal{O}(10^5)$ \citep[][]{2018Natur.553..182M}.
Repeating sources also tend to display stronger spectral depolarization \cite[][]{2022Sci...375.1266F}.}
Together, these suggest that the burst originated in a far less magnetoionically active environment than those of FRB~20121102A and \fastfrb. However, this burst does have a relatively high circular polarization fraction when compared with other non-repeating FRBs ($0.39 \pm 0.01$; \citealt{Scott2025}). \changes{Our limit on the presence of a PRS is a factor of $50$ lower than the observed PRS luminosity for repeating FRBs 20121102A  and 20190520B \cite[][]{clw+17,Niu22}; this also points to a less magnetoionically active environment.
Overall, none of the discussed radio properties of \ourfrb\ (scattering, local DM contribution, rotation measure (RM), and coincident continuum radio source) require the presence of a dense, turbulent, and/or highly magnetized ionized circumburst region like those inferred for FRB~20121102A and FRB~20190520B.}

The presence of this dwarf galaxy within the non-repeating FURBY sample presented in Appendix~\ref{ap:FURBY_DR} 
indicates a \changes{non-negligible} rate 
\changes{(1 of 12 or $\sim 8\%$)}
of dwarf galaxy hosts within the FRB population. 
As shown in Figure~\ref{fig:Lstar}, we note that galaxies as faint as \ourfrb\ would be excluded from our sample by $z\sim0.2$ per the criteria presented in Section~\ref{sec:obs}. We emphasize that this result 
\changes{stands} in contrast to 
the FRB host population \changes{study} presented in \citet{Sharma2024}, \changes{which inferred that the 
dominant FRB channel is biased against low metallicity
stars and hence low mass galaxies. 
We find, however, that the stellar mass of \ourfrb\ 
falls 1 order of magnitude below their estimated
cutoff in host galaxy mass.
If there is a low metallicity bias \citep[see][for a counter argument]{horowicz2025}, it must not preclude
the formation of FRBs in systems like
that hosting \ourfrb.}
\changes{We emphasize that}
in comparison to \cite{Sharma2024} \changes{and other 
previous work}, 
the FURBY sample includes deeper imaging which allows us to identify fainter host galaxies\changes{:} FURBY is limited to $m_R\sim 24$, 0.5 magnitude deeper than what is reported in the \cite{Sharma2024} sample. 
Furthermore, given the PanSTARRS $5\sigma$ point source depth of only $m_r = 23.2$ mag, two of the 12 FURBY host galaxies presented here (corresponding to FRB\,20220610A and FRB\,20220918A) would also
not be detectable using their methods.

As the samples of FRB host galaxies continue to
grow, it will become increasingly important to 
define and account for the selection criteria 
(and resultant biases) that define individual
samples. This is particularly important as one 
attempts to collate multiple surveys to infer 
attributes of the overall population.
With FURBY, we have obtained homogeneous, deep
imaging and spectroscopy with an 8\,m telescope 
to detect even very faint galaxies like \changes{the host of} \ourfrb.
Nevertheless, this sample is also incomplete
\citep[e.g.][]{Marnoch2023} and therefore
biased against galaxies like \ourfrb\ at
high redshift. 
To identify and correct for these effects,
we emphasize the importance of strict selection criteria when pursuing future
surveys of large samples of FRB host galaxies.

\section{Acknowledgments}

A.R.M.\ and R.A.J.\ acknowledge support from the National Science Foundation under grant AST-2206492 and from the Nantucket Maria Mitchell Association. 
A.R.M., A.C.G., J.X.P., W.F., R.A.J., and N.T.\ acknowledge support from NSF grants AST-1911140, AST-1910471 and AST-2206490 as members of the Fast and Fortunate for FRB Follow-up team.
A.C.G., W.F., and the Fong Group at Northwestern gratefully acknowledge support by the NSF under grant Nos. AST-1909358, AST-2206494, AST-2308182 and CAREER grant No. AST-2047919. 
\changes{S.B. is supported by Dutch Research Council (NWO) Veni Fellowship VI.Veni.212.058.}
A.T.D., J.J-S.\changes{,} R.M.S.\changes{, and Y.W.}\ acknowledge support through Australian Research Council Discovery Project DP220102305.
W.F.\ gratefully acknowledges support by the David and Lucile Packard Foundation, the Alfred P.\ Sloan Foundation, and the Research Corporation for Science Advancement through Cottrell Scholar Award \#28284.
M.G.\ is supported through UK STFC Grant ST/Y001117/1. M.G.\ acknowledges support from the Inter-University Institute for Data Intensive Astronomy (IDIA). IDIA is a partnership of the University of Cape Town, the University of Pretoria and the University of the Western Cape. For the purpose of open access, M.G. has applied a Creative Commons Attribution (CC BY) licence to any Author Accepted Manuscript version arising from this submission.
\changes{M.G.\ and C.W.J.\ acknowledge support through Australian Research Council Discovery Project DP210102103.}
R.M.S.\ \changes{and P.A.U.}\ acknowledge support through Australian Research Council Future Fellowship FT190100155.
N.T.\ acknowledges support by FONDECYT grant 1252229.
\changes{CRACO was funded through Australian Research Council Linkage Infrastructure Equipment, and Facilities grant LE210100107.}
This work is based on observations collected at the European
Southern Observatory under ESO programmes 105.204W and 108.21ZF.

This scientific work uses data obtained from Inyarrimanha Ilgari Bundara, the CSIRO Murchison Radio-astronomy Observatory. We acknowledge the Wajarri Yamaji People as the Traditional Owners and native title holders of the Observatory site. CSIRO’s ASKAP radio telescope is part of the Australia Telescope National Facility (\url{https://ror.org/05qajvd42}). Operation of ASKAP is funded by the Australian Government with support from the National Collaborative Research Infrastructure Strategy. ASKAP uses the resources of the Pawsey Supercomputing Research Centre. Establishment of ASKAP, Inyarrimanha Ilgari Bundara, the CSIRO Murchison Radio-astronomy Observatory and the Pawsey Supercomputing Research Centre are initiatives of the Australian Government, with support from the Government of Western Australia and the Science and Industry Endowment Fund.

\vspace{5mm}
\facilities{VLT:Antu (FORS2), VLT:Kueyen (X-Shooter),
VLT:Yepun (HAWK-I)\changes{, ATCA (CABB)}}

\software{
\texttt{Astropy} \citep{astropy}, 
\texttt{Dynesty} \citep{Speagle2020},
\texttt{Extinction} (\url{https://github.com/sncosmo/extinction}),
\texttt{FRBs/FRB} \citep{FRB_Repo}, 
\texttt{Ginga} (\url{https://github.com/ejeschke/ginga}),
\texttt{Linetools} \citep{linetools},
\texttt{Matplotlib} \citep{matplotlib},
\texttt{Numpy} \citep{numpy}, 
\texttt{Pandas} \citep{pandas},
\texttt{pPXF} \citep{ppxf}, 
\texttt{Prospector} \citep{Johnson+21},
\texttt{PypeIt} \citep{pypeit:joss_pub,pypeit:zenodo},
\texttt{Scipy} \citep{scipy}
          }

\appendix

\section{FURBY Spectroscopic Data Release I} \label{ap:FURBY_DR}

\subsection{Spectroscopic Observations} \label{ap:obs_details}

\begin{figure}
    \centering
    \gridline{\fig{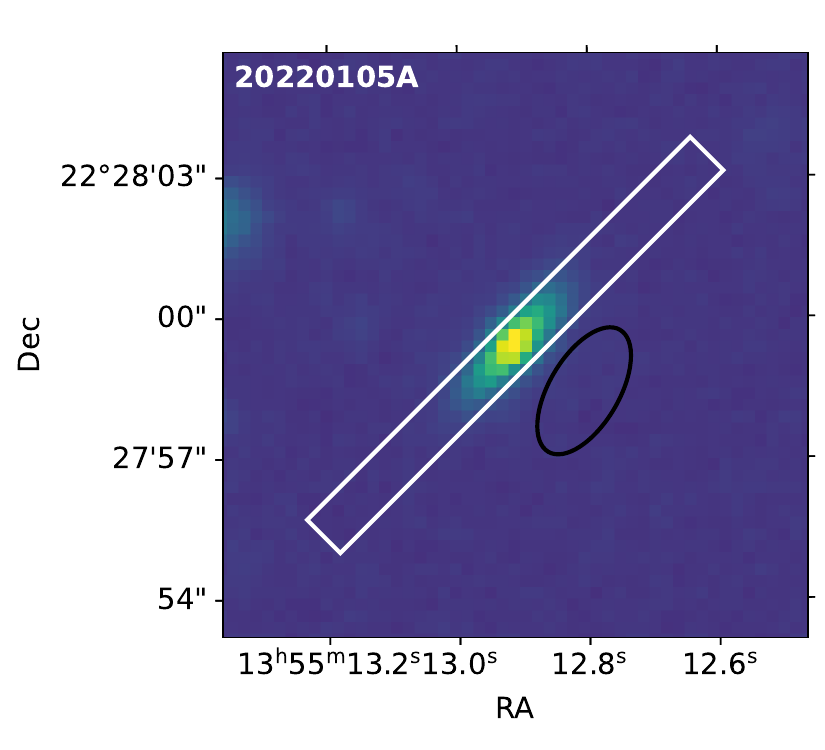}{0.25\textwidth}{}
          \fig{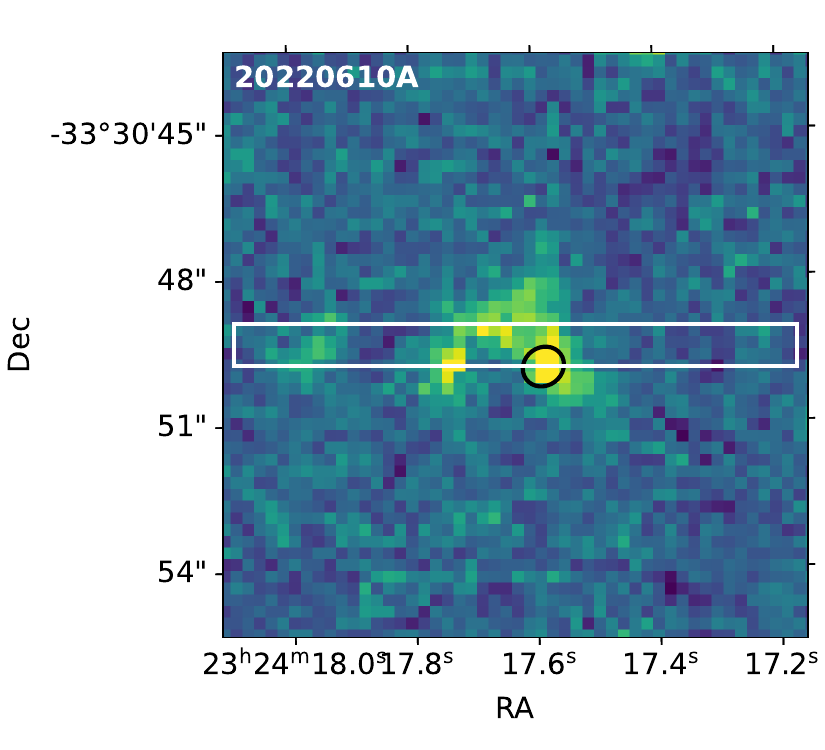}{0.25\textwidth}{}
          \fig{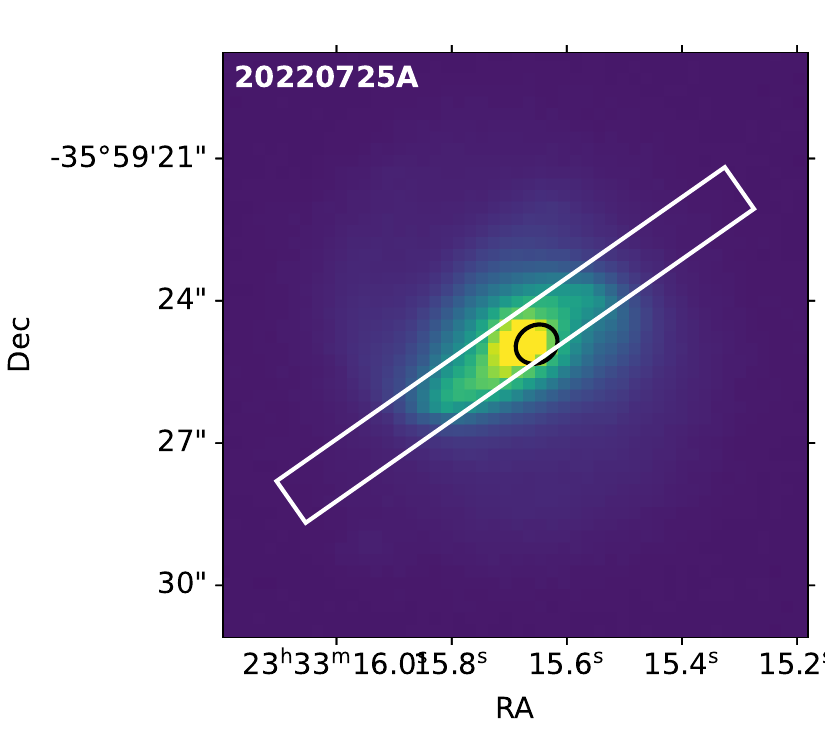}{0.25\textwidth}{}
          \fig{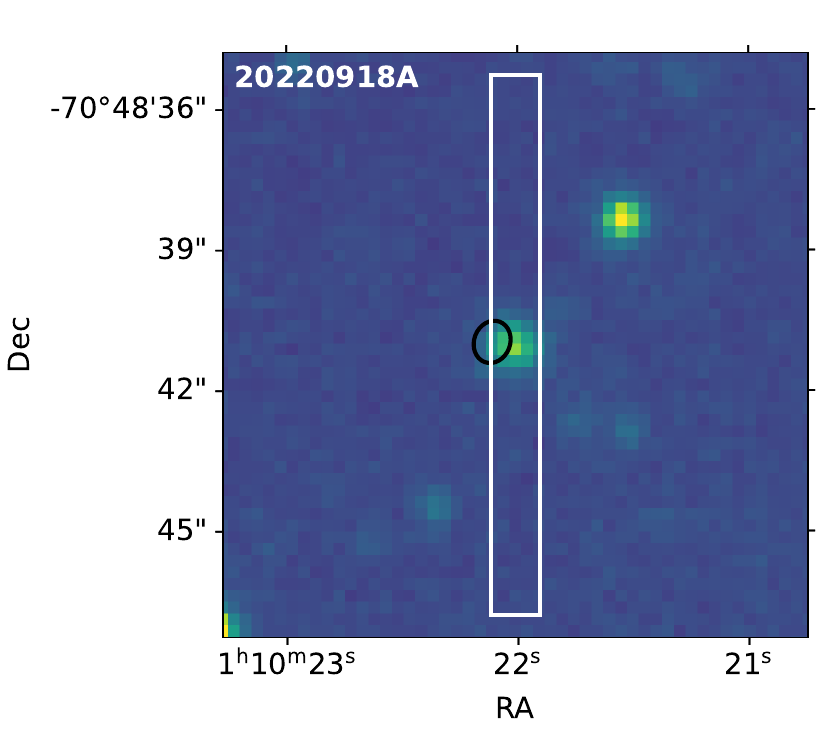}{0.25\textwidth}{}}
    \gridline{\fig{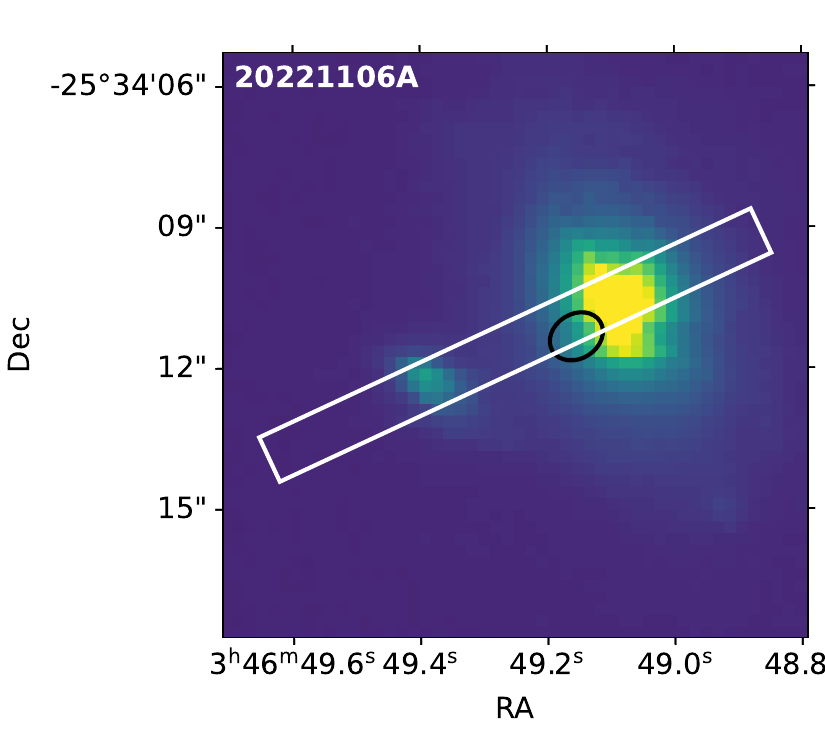}{0.25\textwidth}{}
          \fig{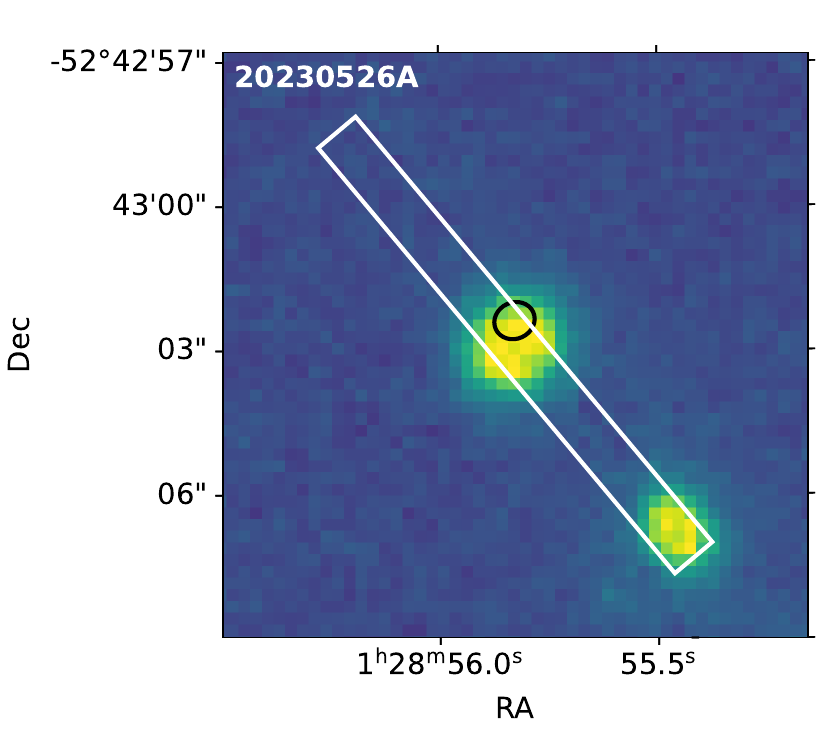}{0.25\textwidth}{}
          \fig{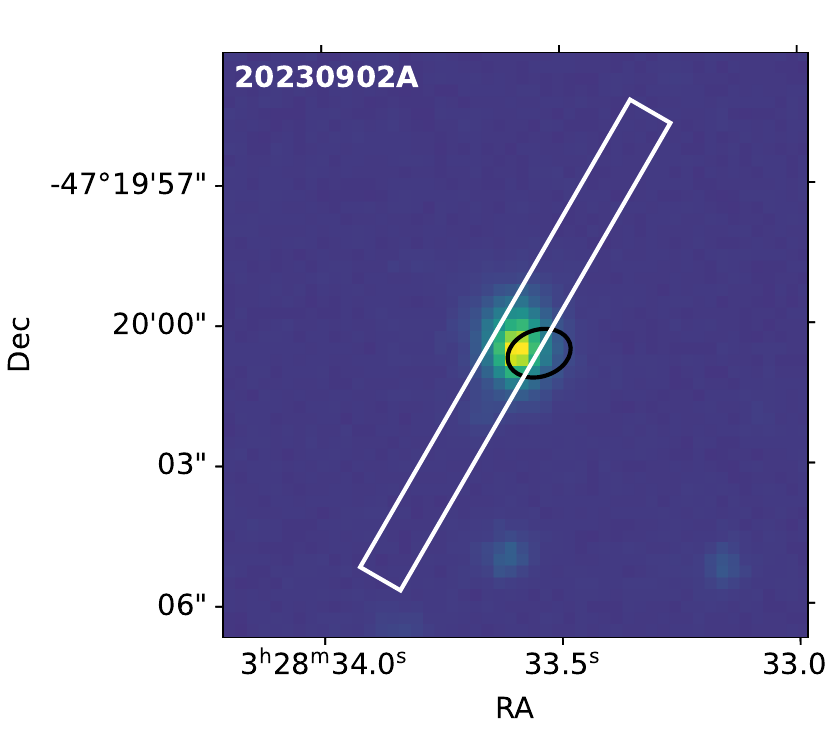}{0.25\textwidth}{}
          \fig{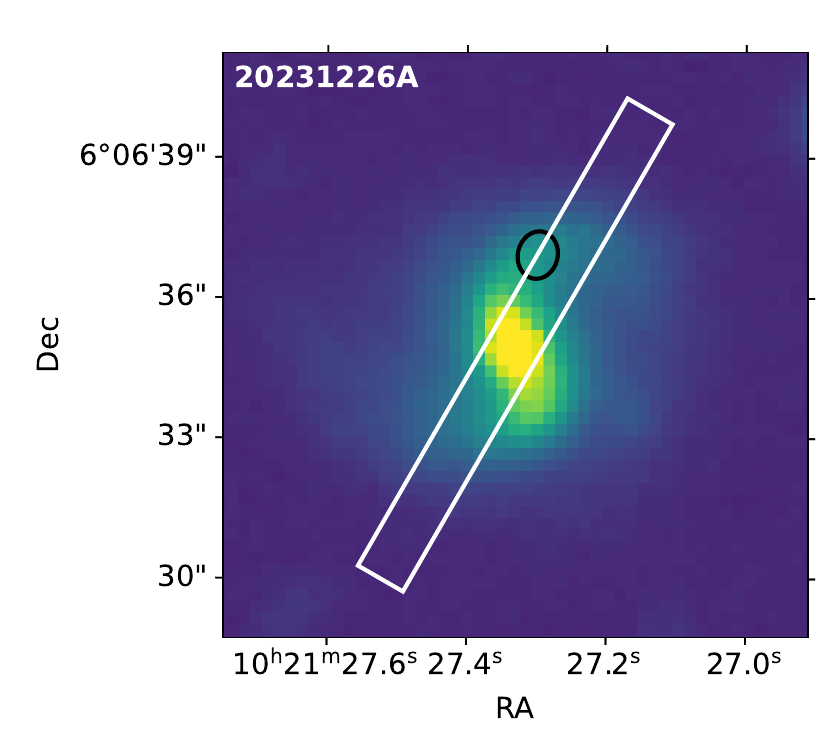}{0.25\textwidth}{}}
    \gridline{\fig{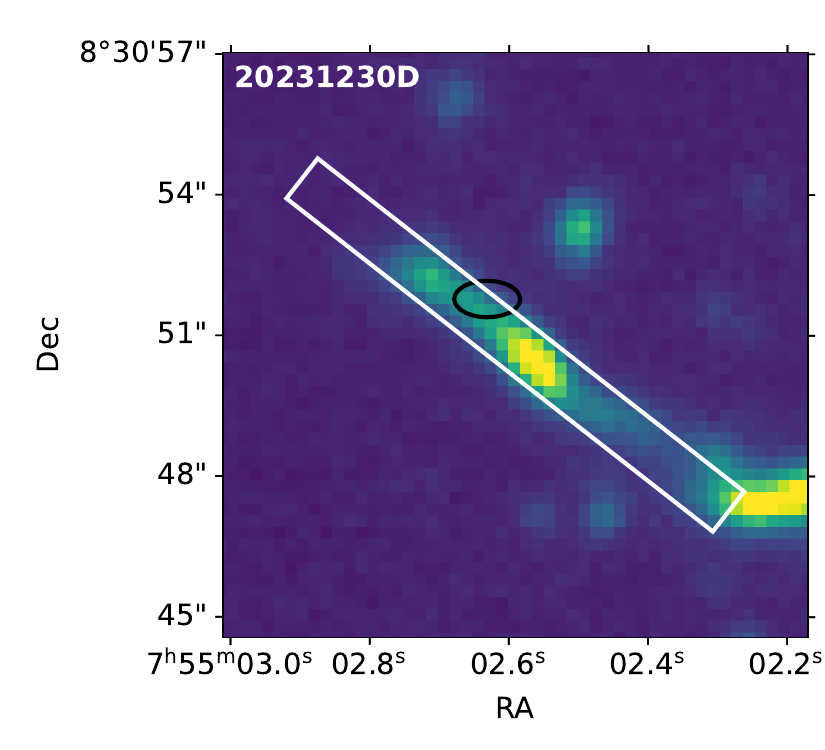}{0.25\textwidth}{}
          \fig{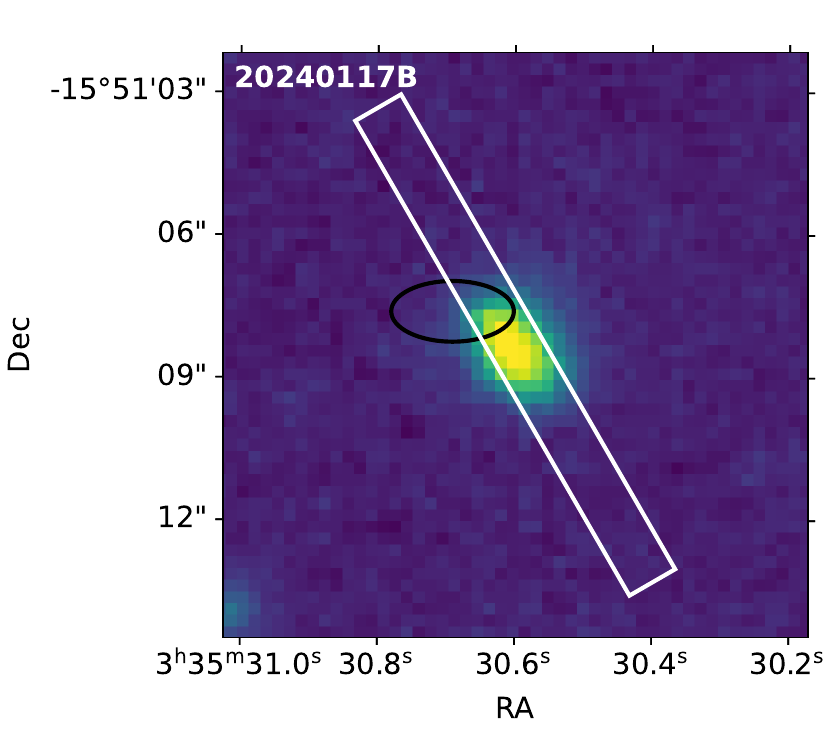}{0.25\textwidth}{}
          \fig{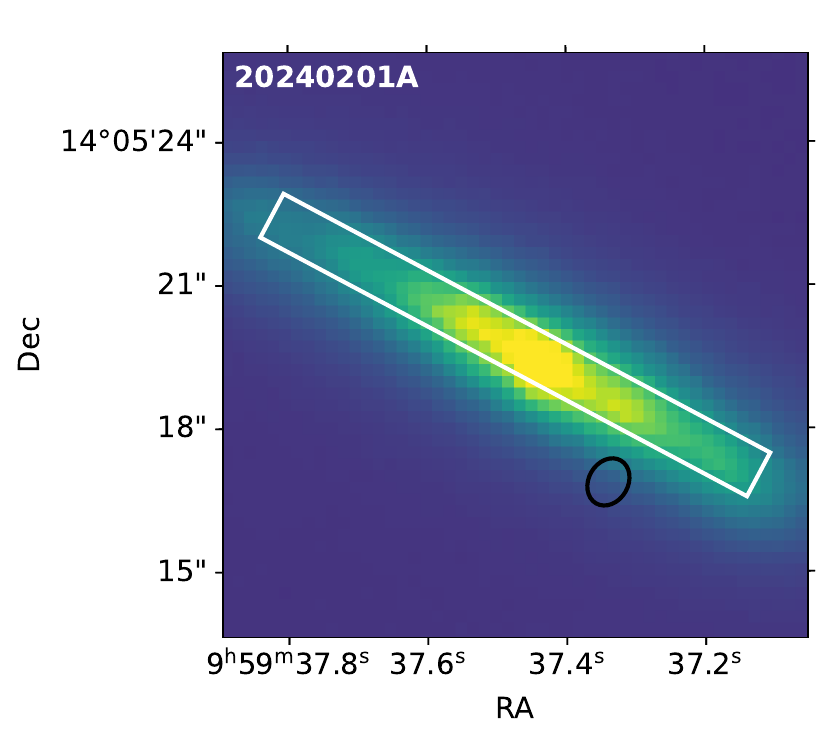}{0.25\textwidth}{}}
    \caption{VLT/FORS2 $R$-band images of all FURBY hosts included in this release. The VLT\changes{/}X\changes{-}Shooter slit positions are shown as white boxes and FRB localizations as black ellipses.}
    \label{fig:slits}
\end{figure}

We present a sample of 12 FRB host galaxies associated with FURBY candidates, of which one is the focus of this work (\ourfrb). $R$-band images of each host galaxy, including the FRB localization and \xshootslit~position, are shown in Figure~\ref{fig:slits}. The details of this host galaxy sample, as well as exposure times and other observing/analysis specifics, are given in Table \ref{tab:observations}.

Most of the FURBY host galaxies exhibit clear nebular emission lines, and in some cases a strong stellar continuum, which allowed for straightforward extraction in the \texttt{PypeIt} workflow. The spectra were then co-added in one dimension to produce the final data product (see Figures \ref{fig:spectra}-\ref{fig:spectra_contd}). Dimmer hosts that did not exhibit clear gas emission in the individual science images were instead first co-added in two dimensions using \texttt{PypeIt} before the one-dimensional host galaxy spectrum was extracted. The co-addition method used for each host is given in the final column of Table \ref{tab:observations}. Even after co-addition, the FRB\,20220610A host did not contain sufficiently bright emission for automatic extraction in \texttt{PypeIt} and was therefore extracted manually based on visual identification of the [\ion{O}{2}] doublet \citep[for further discussion see][]{Gordon2024}. All 1D spectra extracted as part of this work are publicly accessible via doi:\href{https://doi.org/10.5061/dryad.d2547d8hh}{10.5061/dryad.d2547d8hh}.

\begin{deluxetable}{lccccrrrrrcc}
\rotate
\tablecaption{Spectroscopic observation information for all given FURBY hosts}
\tablehead{ &  &  &  &  & PATH & Slit PA & UVB Exp. & VIS Exp. & NIR Exp. &  & Coadded \\
FRB & FRB RA & FRB Decl. & Host RA & Host Decl. & Prob. (\%) & (deg) & Time (s) & Time (s) & Time (s) & Mode & in 2D?}
\startdata
20220105A & 13:55:12.81 & +22:27:58.4 & 13:55:12.91 & +22:27:59.41 & 99.98 & 135 & 2400 & 2400  & 2400 & STARE  & No  \\
20220610A & 23:24:17.58 & $-$33:30:49.9 & 23:24:17.63 & $-$33:30:49.41 & 100 & 90 & 7404 & 7548 & 7200 & NOD & No  \\
20220725A & 23:33:15.65 & $-$35:59:24.9 & 23:33:15.69 & $-$35:59:24.94 & 100 & 125 & 2400 & 2400  & 2400 & STARE  & No  \\
20220918A & 01:10:22.11 & $-$70:48:41.0 & 01:10:22.01 & $-$70:48:41.03 & 100 & 0 & 4800 & 4800  & 4800 & STARE  & Yes \\
20221106A & 03:46:49.15 & $-$25:34:11.3 & 03:46:49.07 & $-$25:34:10.65 & 99.98 & 115 & 2400 & 2400 & 2400 & STARE  & No  \\
20230526A & 01:28:55.83 & $-$52:43:02.4 & 01:28:55.83 & $-$52:43:02.90 & 99.99 & 40 & 2400 & 2400  & 2400 & STARE  & No  \\
20230708A & 20:12:27.73 & $-$55:21:22.6 & 20:12:27.73 & $-$55:21:22.72 & 100 & 0 & 2400 & 2400  & 2400 & STARE  & Yes \\
20230902A & 03:28:33.55 & $-$47:20:00.6 & 03:28:33.60 & $-$47:20:00.43 & 100 & 150 & 2400 & 2400 & 2400 & STARE  & No  \\
20231226A & 10:21:27.30 & +06:06:36.9 & 10:21:27.33 & +06:06:35.01 & 100 & 150 & 1200 & 1200  & 1200 & STARE  & No  \\
20231230D & 07:55:02.63 & +08:30:51.8 & 07:55:02.59 & +08:30:50.82 & 99.94 & 52 & 2400 & 2400  & 2400 & STARE  & Yes \\
20240117B & 03:35:30.69 & $-$15:51:07.6 & 03:35:30.60 & $-$15:51:08.30 & 100 & 30 & 7404 & 7548  & 7200 & NOD & Yes \\
20240201A & 09:59:37.34 & +14:05:16.9 & 09:59:37.47 & +14:05:19.76 & 99.98 & 62 & 300  & 329   & 300 & STARE  & No \\
\enddata
\tablecomments{Interferometric FRB positions are listed without error. See \citet{shannon2025} for localization details of all FRBs except 20231230D and 20240117A, which are presented in \citet{Gordon+25}.}
\label{tab:observations}
\end{deluxetable}

\subsection{Fluxes} \label{ap:fluxes}

We measured nebular emission line fluxes for all FURBY host galaxies using the pPXF spectral fitting package, with results presented in Table~\ref{tab:fluxes}. The measured fluxes include key diagnostic lines: \halpha, \hbeta, [\ion{N}{2}]$\lambda6583$, and [\ion{O}{3}]$\lambda5007$, though we note that these represent only a subset of the emission lines detected across the sample. Several hosts also show additional features including [\ion{O}{2}]$\lambda\lambda3726,3729$ and in some cases higher-order hydrogen series lines such as Paschen and Brackett transitions. The emission line measurements were extracted from the final reduced one-dimensional spectra, with uncertainties determined from the spectral error arrays propagated through the fitting process.

For the faintest hosts in the sample (FRB~20220610A, FRB~20220918A), the measured emission lines fall below our detection thresholds, resulting in upper limits only. 
The flux measurements in Table~\ref{tab:fluxes} have been corrected for Galactic extinction using the \cite{Fitzpatrick:2007} law but have not been corrected for host galaxy internal extinction. These line flux measurements serve as the foundation for our metallicity estimates and star formation rate calculations presented in the main analysis.

\begin{figure}
    \centering
    \gridline{\fig{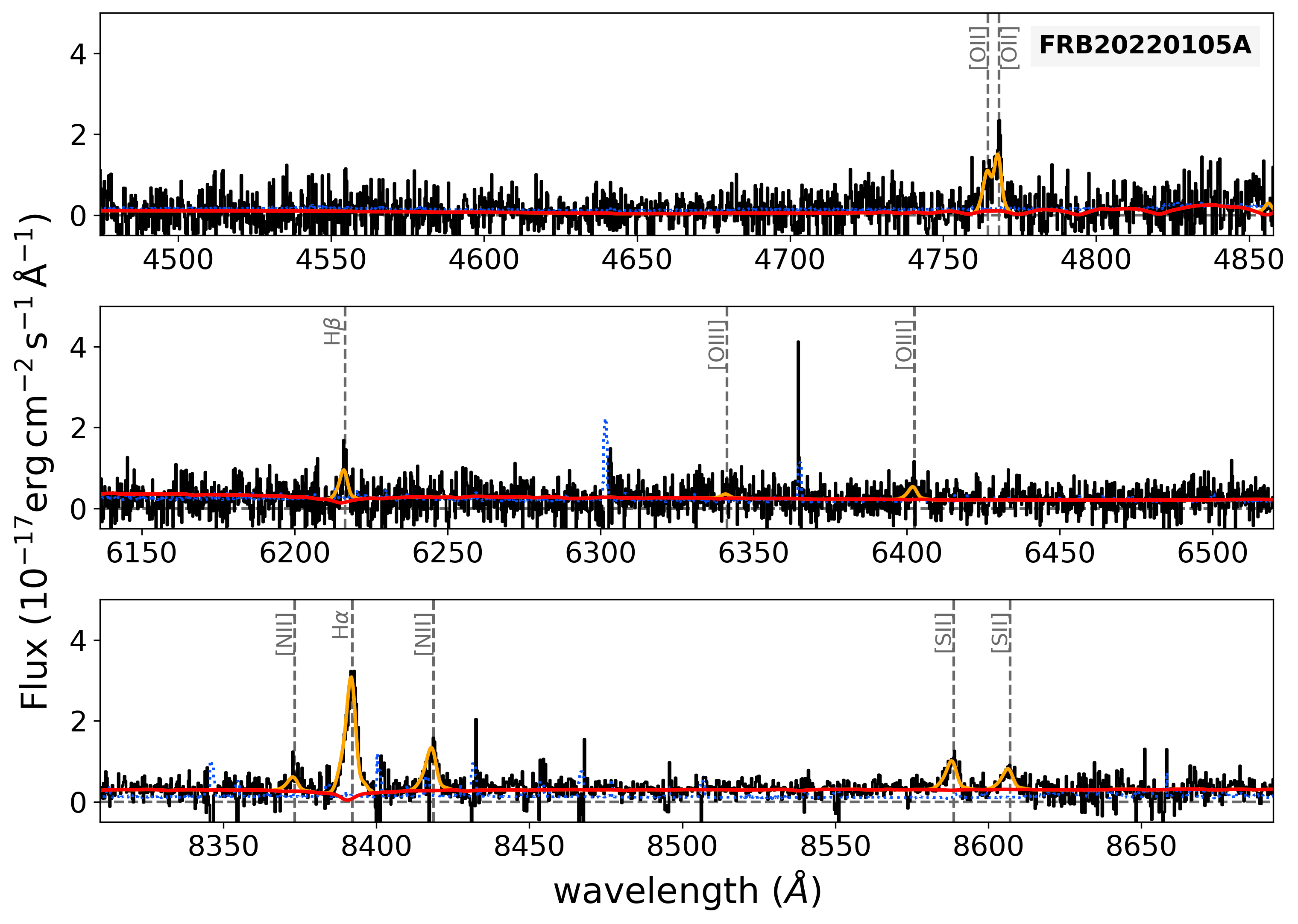}{0.49\textwidth}{}
          \fig{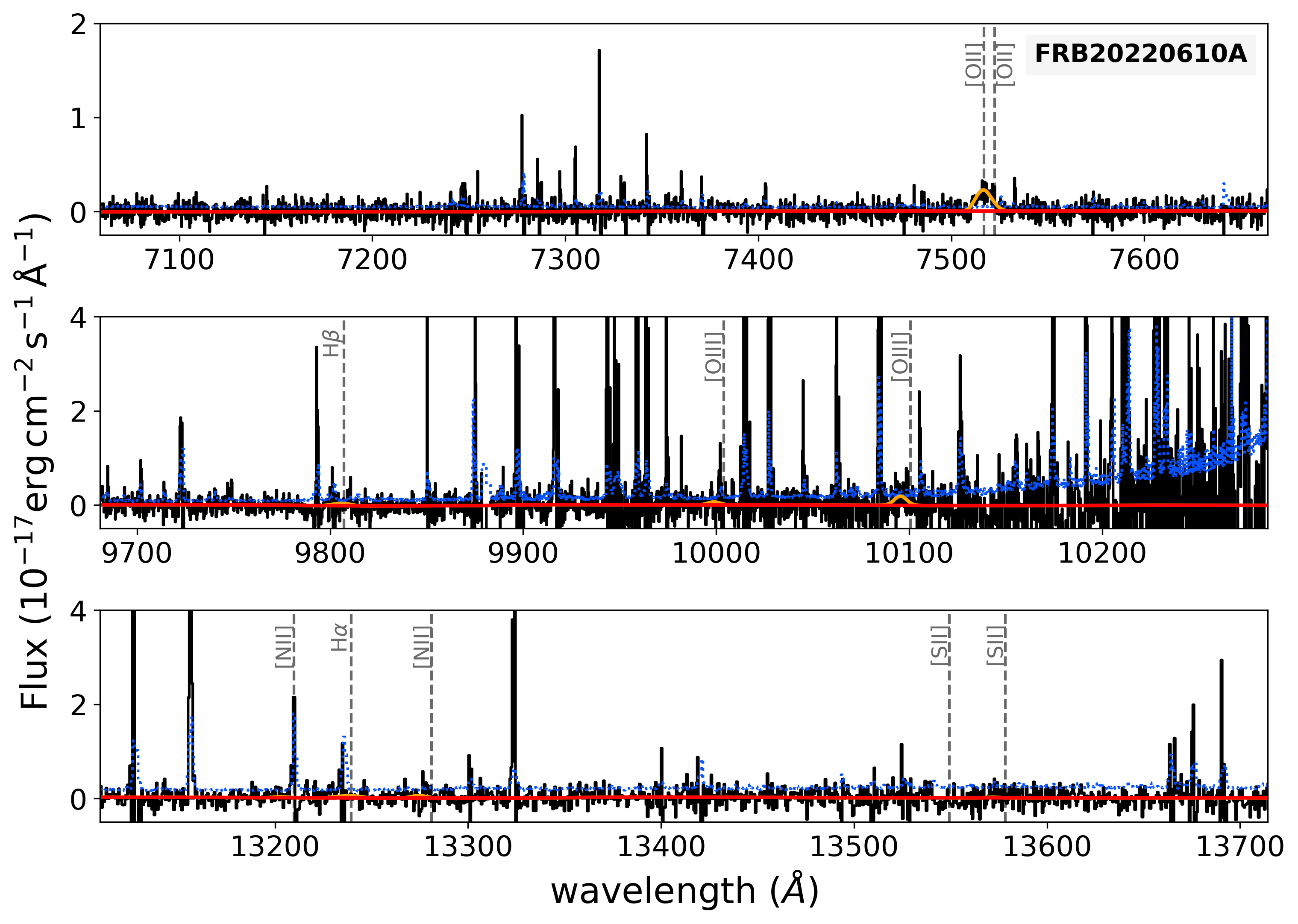}{0.49\textwidth}{}}
    \gridline{\fig{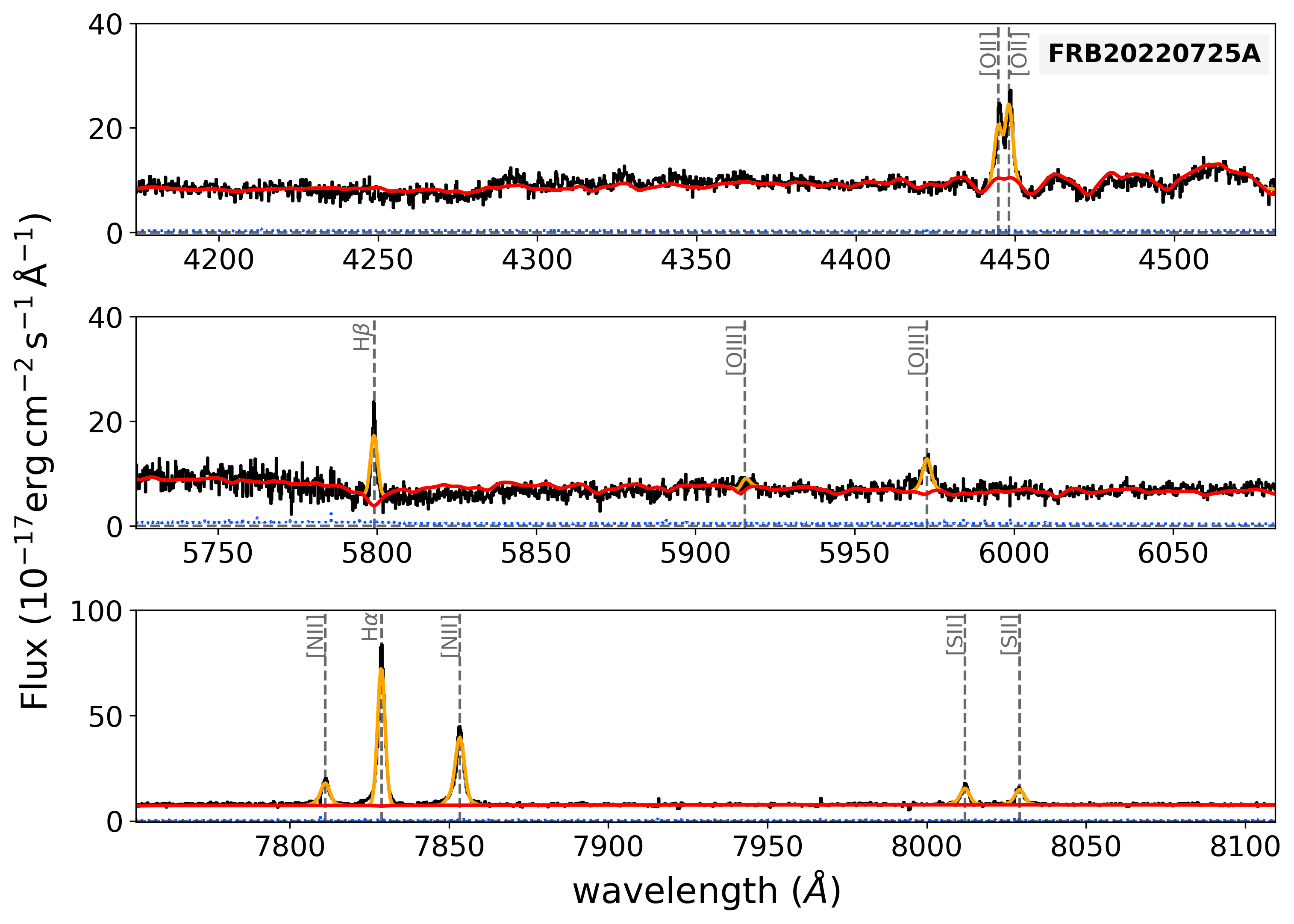}{0.49\textwidth}{}
          \fig{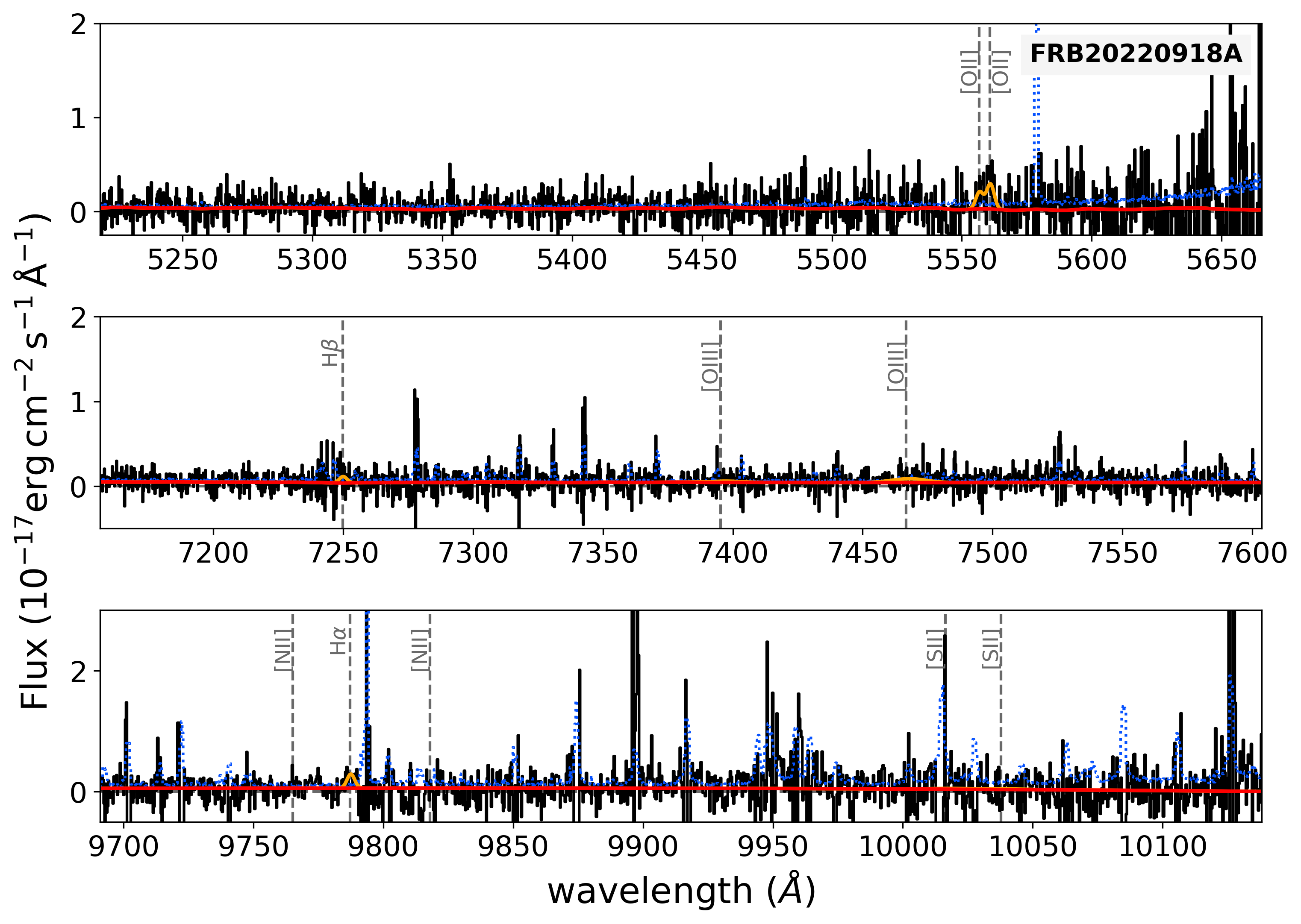}{0.49\textwidth}{}}
    \gridline{\fig{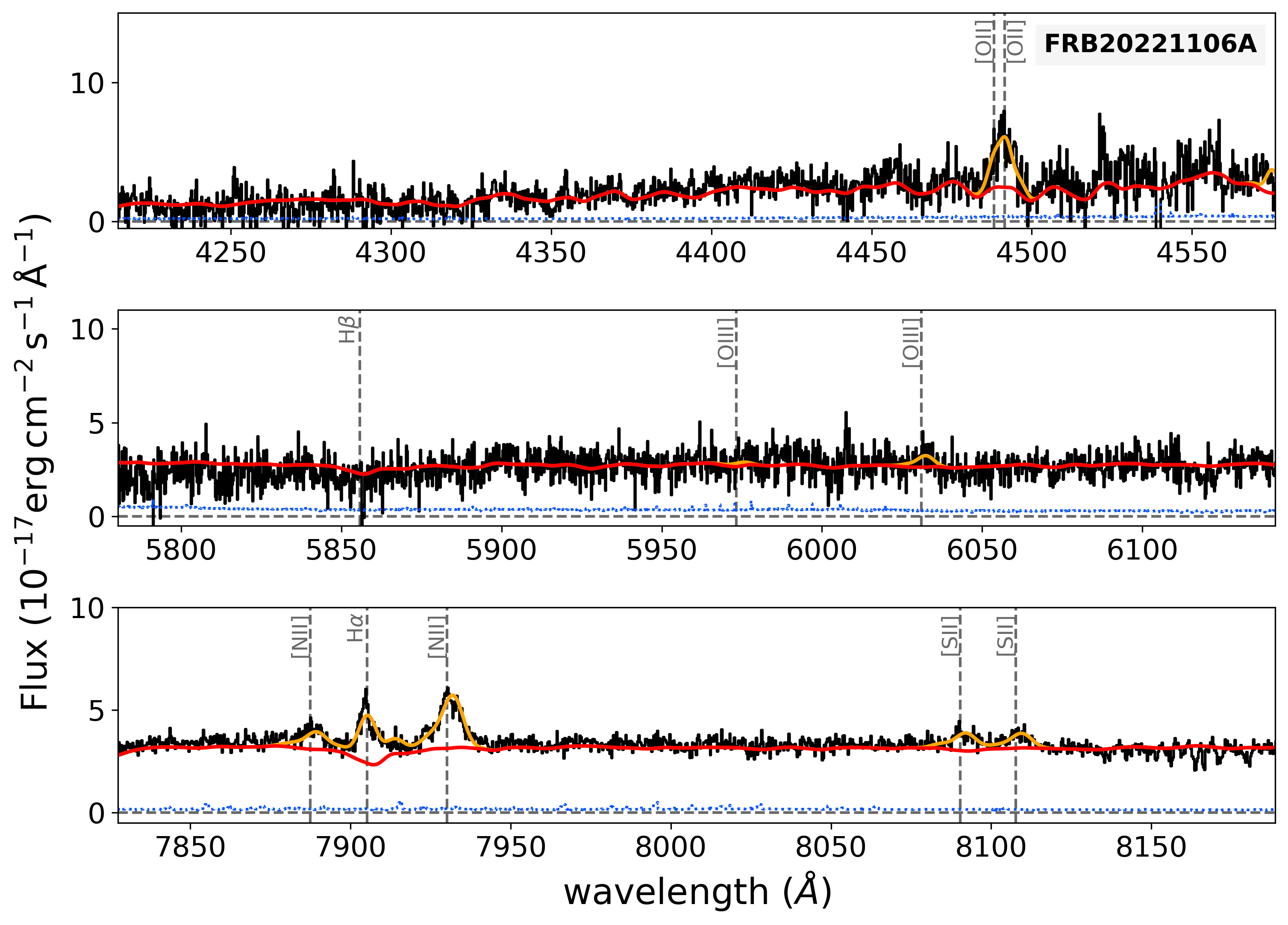}{0.49\textwidth}{}
          \fig{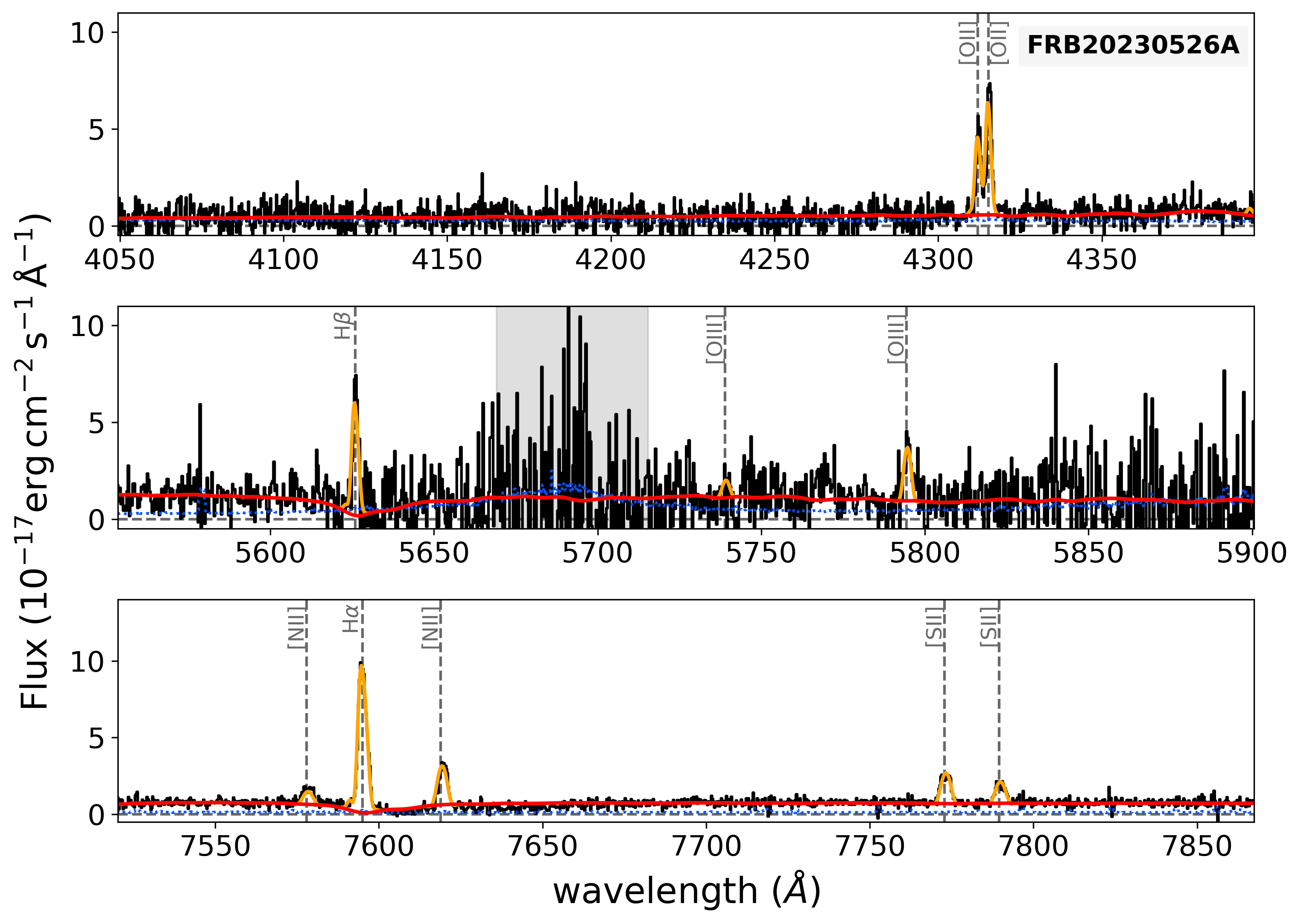}{0.49\textwidth}{}}
    \caption{Selected emission features for each FRB in the FURBY sample. The pPXF gas emission fit is shown in orange, while the stellar continuum fit is shown in red. Spectral error is shown in blue. A gray shaded region indicates part of the spectrum that was excluded from the fit.}
    \label{fig:spectra}
\end{figure}
\begin{figure}
    \centering
        \gridline{\fig{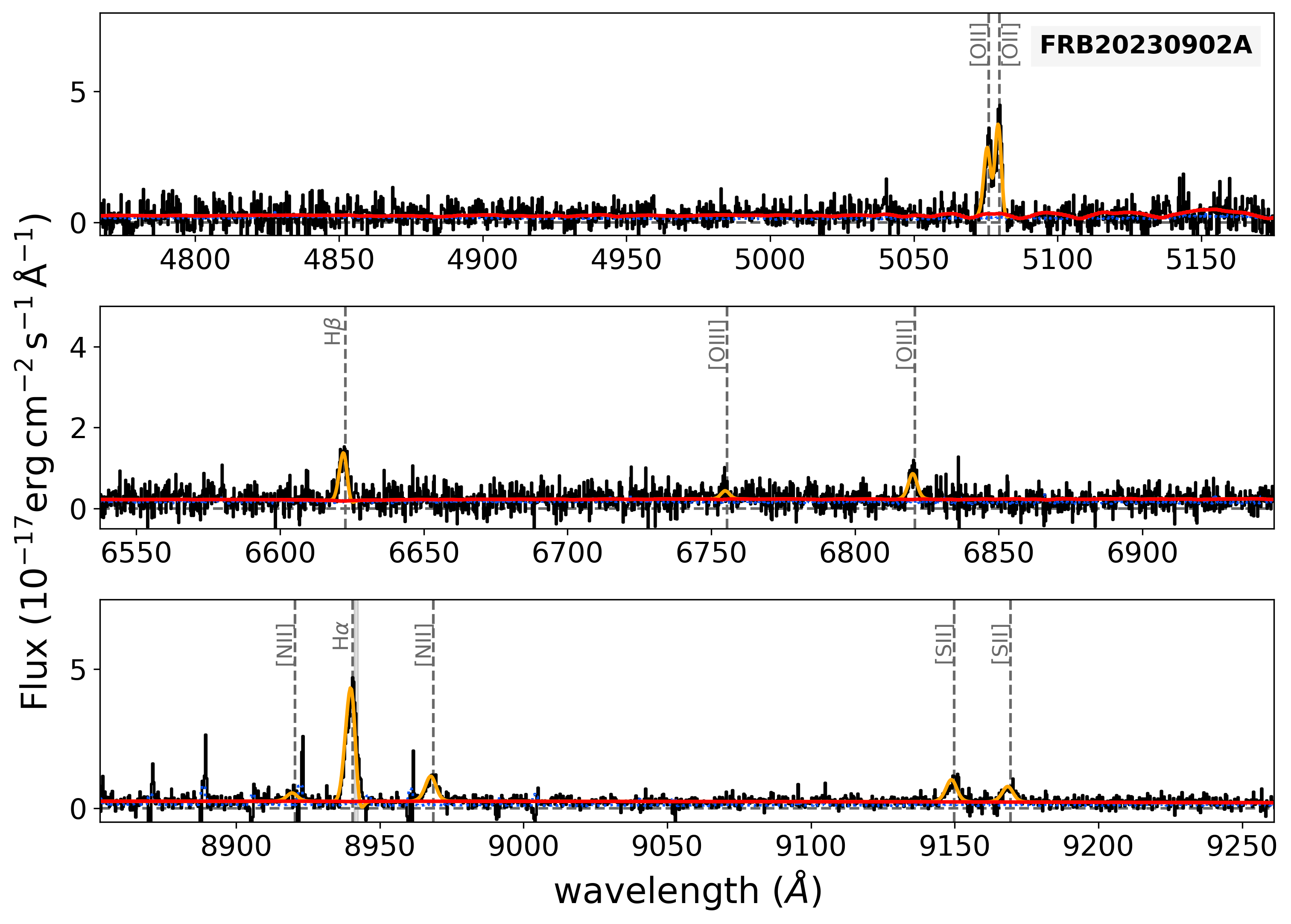}{0.49\textwidth}{}
          \fig{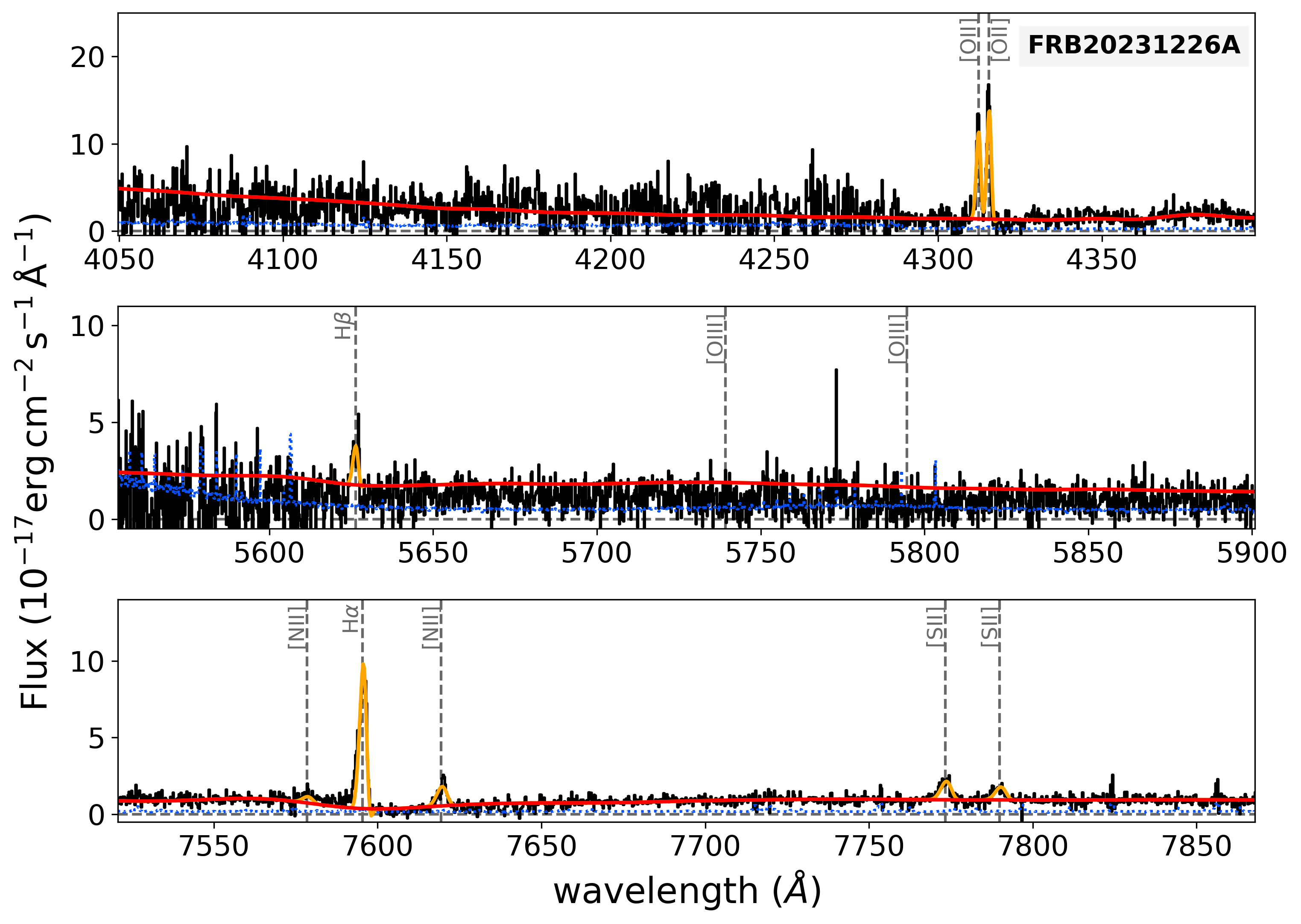}{0.49\textwidth}{}}
        \gridline{\fig{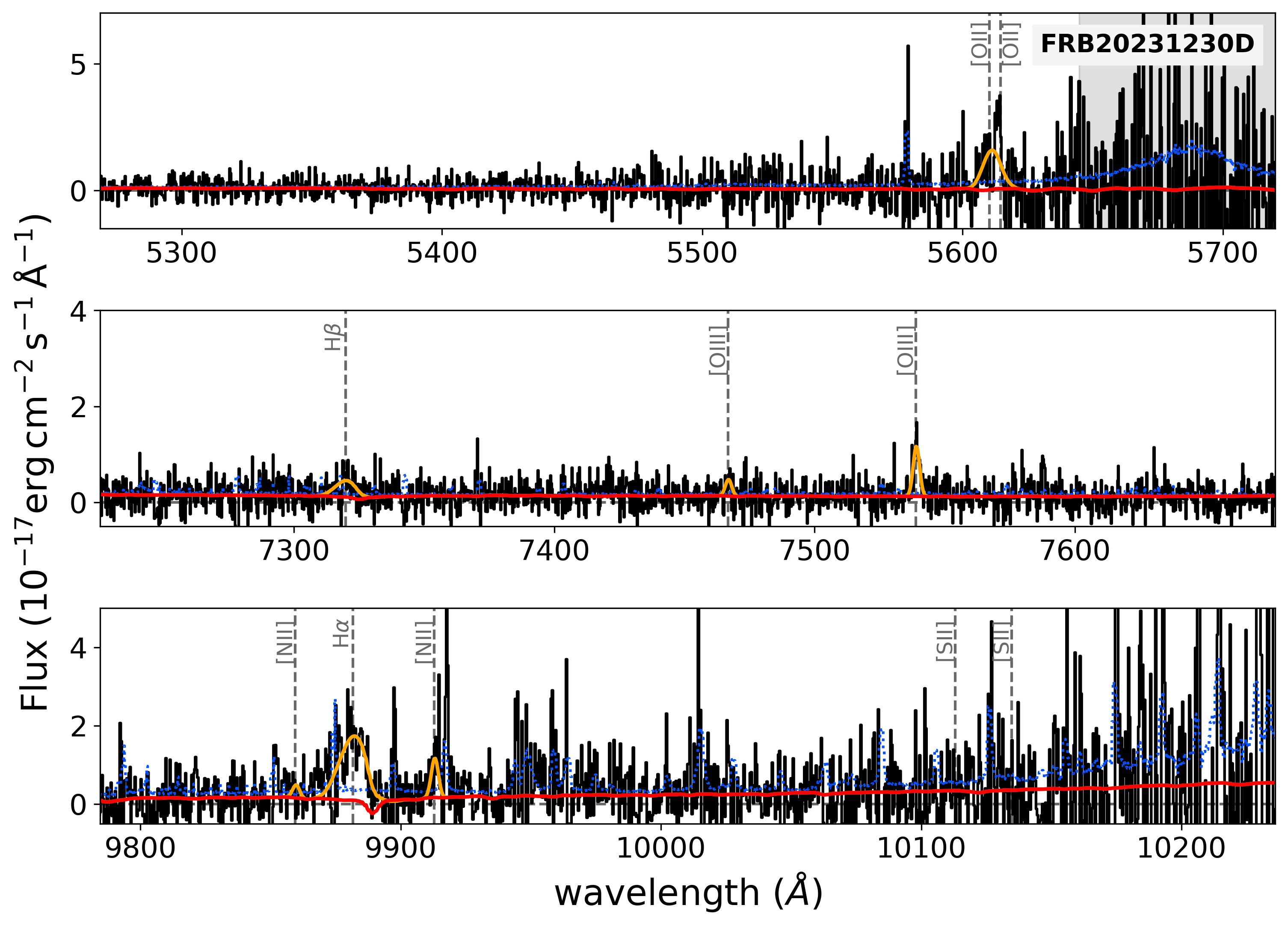}{0.49\textwidth}{}
          \fig{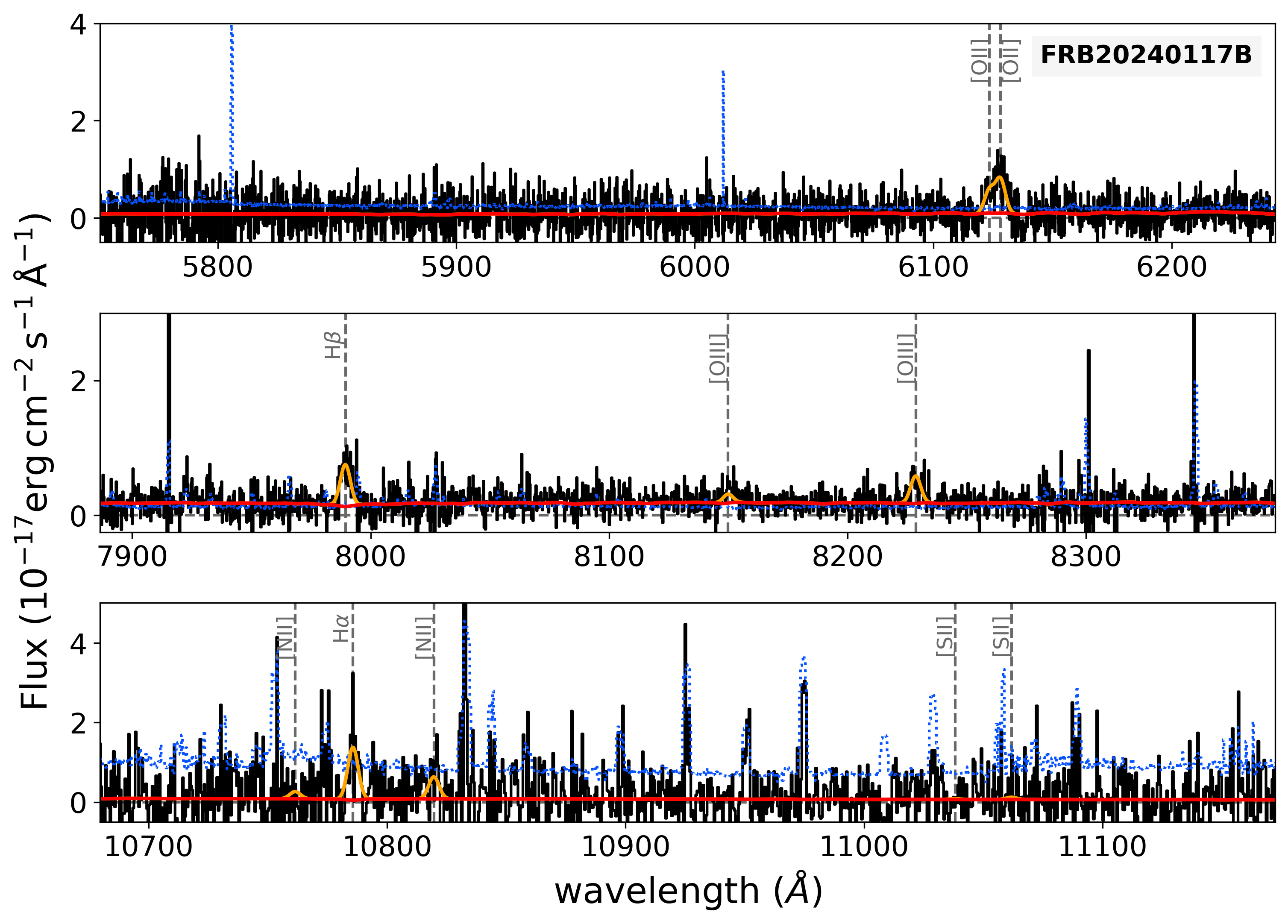}{0.49\textwidth}{}}
    \gridline{\fig{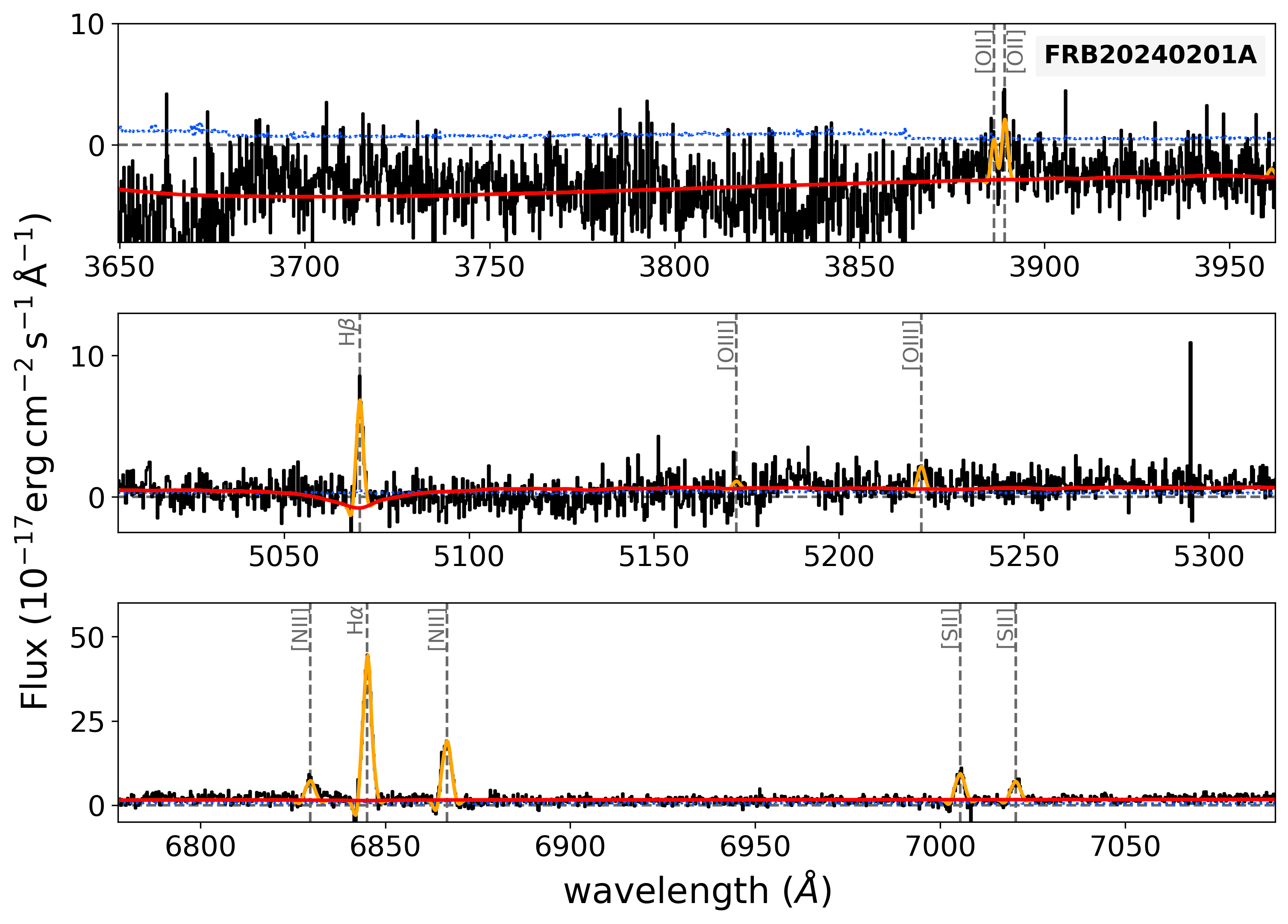}{0.49\textwidth}{}}
    \caption{(Continued) selected emission features for each FRB in the FURBY sample.}
    \label{fig:spectra_contd}
\end{figure}

\input{Table2}

\subsection{Sample Demographics} \label{ap:trends}

To supplement the analysis presented in Section \ref{sec:results}, we compute an updated redshift, star formation rate, and metallicity for each FURBY host using the flux measurements as described in Appendix \ref{ap:fluxes}. These galaxy properties, along with the observed $R$-band magnitude and a by-eye assessment of the Balmer absorption as modeled in pPXF, are given in Table \ref{tab:host_properties}.

\input{Table3}

\bibliography{bibfile}{}
\bibliographystyle{aasjournal}

\end{document}

%% file: affiliations.tex
\newcommand{\MMO}{\affiliation{Maria Mitchell Observatory,
Nantucket, MA 02554, USA}}
\newcommand{\AEI}{\affiliation{Max Planck Institute for Gravitational Physics (Albert Einstein Institute), 14476 Potsdam, Germany}}
\newcommand{\Glasgow}{\affiliation{School of Physics and Astronomy, University of Glasgow, Glasgow, G12 8QQ, United Kingdom}}
\newcommand{\MacquarieMathPhys}{\affiliation{School of Mathematical and Physical Sciences, Macquarie University, Sydney, NSW 2109, Australia}}
\newcommand{\MacquarieSpace}{\affiliation{Astrophysics and Space Technologies Research Centre, Macquarie University, Sydney, NSW 2109, Australia}}
\newcommand{\CIERA}{\affiliation{Center for Interdisciplinary Exploration and Research in Astrophysics (CIERA), Northwestern University, Evanston, IL 60201, USA}}
\newcommand{\NU}{\affiliation{Department of Physics and Astronomy, Northwestern University, Evanston, IL 60208, USA}}
\newcommand{\NAOJ}{\affiliation{Division of Science, National Astronomical Observatory of Japan, 2-21-1 Osawa, Mitaka, Tokyo, 181-8588, Japan}}
\newcommand{\IPMU}{\affiliation{Kavli Institute for the Physics and Mathematics of the Universe (Kavli IPMU), 5-1-5 Kashiwanoha, Kashiwa, 277-8583, Japan}}
\newcommand{\UCSC}{\affiliation{Department of Astronomy and Astrophysics, University of California, Santa Cruz, CA 95064, USA}}
\newcommand{\Curtin}{\affiliation{International Centre for Radio Astronomy Research, Curtin University, Bentley, WA 6102, Australia}}
\newcommand{\CSIRO}{\affiliation{Australia Telescope National Facility, CSIRO, Space and Astronomy, PO Box 76, Epping, NSW 1710, Australia}}
\newcommand{\SKAO}{\affiliation{SKA Observatory (SKAO), Science Operations Centre, CSIRO ARRC, Kensington WA 6151, Australia}}
\newcommand{\Swinburne}{\affiliation{Centre for Astrophysics and Supercomputing, Swinburne University of Technology, Hawthorn, VIC 3122, Australia}}
\newcommand{\Sydney}{\affiliation{Sydney Institute for Astronomy, School of Physics, The University of Sydney, NSW 2006, Australia}}
\newcommand{\Edinburgh}{\affiliation{Institute for Astronomy, University of Edinburgh, Royal Observatory, Edinburgh, EH9 3HJ, United Kingdom}}
\newcommand{\CapeTown}{\affiliation{Inter-University Institute for Data Intensive Astronomy, Department of Astronomy, University of Cape Town, Cape Town, South Africa}}
\newcommand{\Valparaiso}{\affiliation{Instituto de F\'isica, Pontificia Universidad Cat\'olica de Valpara\'iso, Casilla 4059, Valpara\'iso, Chile}}
\newcommand{\HumboldtCA}{\affiliation{Department of Physics \& Astronomy, California State Polytechnic University, Humboldt, Arcata, California 95521, USA}}
\newcommand{\OzGrav}{\affiliation{OzGrav, the ARC Centre of Excellence for Gravitational Wave Discovery, Swinburne University of Technology, Hawthorn, VIC 3122, Australia}}

%% file: Table2.tex
\begin{deluxetable}{lcccc}
\centering
\tablecaption{Measured nebular emission line fluxes for each FURBY host galaxy}
\tablehead{FRB     & H$\alpha$  & H$\beta$   & [\ion{N}{2}] $\lambda6583$ & [\ion{O}{3}] $\lambda5007$ }
\decimals
\startdata
20220105A & $35.10\pm0.66$ & $10.2\pm1.2$ & $12.47\pm0.68$ & $<4.90$\\
20220610A & $<2.6$ & $<2.17$ & $<2.0$ & $<5.3$\\
20220725A & $658.7\pm1.8$ & $202.0\pm3.4$ & $432.6\pm1.9$ & $107.9\pm2.4$\\
20220918A & $<2.20$ & $<1.11$ & $<1.0$ & $<2.63$\\
20221106A & $79.2\pm1.1$ & $9.4\pm2.3$ & $81.3\pm1.2$ & $27.3\pm2.2$\\
20230526A & $88.01\pm0.71$ & $56.2\pm2.0$ & $22.67\pm0.52$ & $25.5\pm1.5$\\
20230708A & $14.02\pm0.49$ & $10.96\pm0.60$ & $<1.37$ & $44.2\pm1.1$\\
20230902A & $36.74\pm0.58$ & $11.76\pm0.70$ & $10.06\pm0.45$ & $7.65\pm0.65$\\
20231226A & $88.0\pm1.2$ & $30.2\pm2.7$ & $21.57\pm0.92$ & $<3.0$\\
20231230D & $56.4\pm2.6$ & $<13.6$ & $8.5\pm1.2$ & $10.87\pm0.65$\\
20240117B & $<21.8$ & $10.68\pm0.69$ & $<10.8$ & $6.80\pm0.56$\\
20240201A & $331.9\pm3.0$ & $64.6\pm1.3$ & $124.9\pm2.3$ & $12.12\pm0.93$
\enddata
\tablecomments{All fluxes are shown in units of $10^{-17}$ erg s$^{-1}$ cm$^{-2}$.}
\label{tab:fluxes}
\end{deluxetable}

%% file: Table3.tex
\begin{deluxetable}{llllDl}
\tablecaption{FURBY host properties}
\tablehead{FRB & $z$ & $m_R$  & Strong Balmer Absorption? & \multicolumn2l{SFR (M$_\odot $~yr$^{-1}$)} & Metallicity [12 + log(O/H)]}
\decimals
\startdata
20220105A & $0.2784$ & $21.270\pm0.005$ & No & $0.425\pm0.008$ & $8.5\pm0.\changes{1}^{\rm a}$ \\
20220610A & $1.017$ & $23.68\pm0.04$ & -- & $<0.3$ & -- \\
20220725A & $0.1926$ & $17.806\pm0.004$ & No & $3.97\pm0.01$ & $^{\rm b}$ \\
20220918A & $0.491$ & $23.58\pm0.02$ & -- & $<0.09$ & -- \\
20221106A & $0.2043$ & $18.322\pm0.009$ & Yes & $0.535\pm0.008$ & $^{\rm b}$ \\
20230526A & $0.1570$ & $21.03\pm0.01$ & No & $0.359\pm0.003$ & $8.\changes{5}\pm0.\changes{1}^{\rm c}$ \\
20230708A & $0.1050$ & $22.53\pm0.02$ & -- & $0.0262\pm0.0009$ & $(\changes{8.0}-8.3)^{\rm a,c}$ \\
20230902A & $0.3619$ & $21.491\pm0.006$ & No & $0.72\pm0.01$ & $8.\changes{5}\pm0.\changes{1}^{\rm c}$ \\
20231226A & $0.1570$ & $18.942\pm0.006$ & Yes & $0.359\pm0.005$ & $8.\changes{5}\pm0.\changes{1}^{\rm a}$ \\
20231230D & $0.505$ & $20.949\pm0.006$ & -- & $2.00\pm0.09$ & $8.\changes{4}\pm0.\changes{2}^{\rm c}$ \\
20240117B & $0.643$ & $22.10\pm0.01$ & -- & $<1.2$ & -- \\
20240201A & $0.0427$ & $16.91\pm0.01$ & No & $0.106\pm0.001$ & $8.6\pm0.\changes{1}^{\rm c}$ \\
\enddata
\tablecomments{$R$-band magnitudes have been corrected for Galactic extinction using the \texttt{Extinction} Python package (\url{https://github.com/sncosmo/extinction}). Dashes indicate insufficient emission detected to compute. Metallicities computed using calibrators given in \citet{O3N2}. 
\newline $^{\rm a}$ Computed using N2. 
\newline $^{\rm b}$ Excluded due to AGN classification (based on Figure~\ref{fig:bpt}). 
\newline $^{\rm c}$ Computed using O3N2.}
\label{tab:host_properties}
\end{deluxetable}